\documentclass{aa}   
\begin{document}
\newcommand{\lesssim}{\stackrel{<}{\sim}}
\newcommand{\gtrsim}{\stackrel{>}{\sim}}
\newcommand{\be}{\vspace{1mm} \begin{equation}} 
\newcommand{\ee}{\end{equation} \vspace{1mm}} 
\newcommand{\der}[2]{\frac{d{#1}}{d{#2}}} 
\newcommand{\dertext}[2]{{d{#1}}/{d{#2}}} 
\newcommand{\pd}[2]{\frac{{\partial}{#1}}{{\partial}{#2}}} 
\newcommand{\pdtext}[2]{{{\partial}{#1}}/{{\partial}{#2}}} 
\newcommand{\func}[1]{\mathrm{#1}} 
\newcommand{\rot}[1]{\mathrm{rot}{#1}} 
\newcommand{\dvrg}[1]{\mathrm{div}{#1}} 
\newcommand{\en}{\varepsilon} 
\newcommand{\nlb}{{\nolinebreak}} 
\newcommand{\kap}{\symbol{"1A}} 
\newcommand{\sR}{{\mbox{\tiny R}}}
\newcommand{\sN}{{\mbox{\tiny N}}}
\title{Eigenoscillations of the differentially rotating Sun: II.
Generalization of Laplace's tidal equation}
\author{N.S. Dzhalilov\inst{1,}\inst{2} \and J. Staude\inst{2}} 
\offprints{J.Staude} 
\institute{Institute of Terrestrial Magnetism,
Ionosphere and Radio Wave Propagation of the Russian Academy of Sciences,\\
Troitsk City, Moscow Region, 142190 Russia; 
E--Mail: namig@izmiran.rssi.ru
\and Astrophysikalisches Institut Potsdam, Sonnenobservatorium Einsteinturm,  
14473 Potsdam, Germany;\\
 E--Mail: jstaude@aip.de}
\date{Received\ \ \ \ \ \ \ \ \ ; accepted\ \ \ \ \ \ \ \ }

\abstract{
The general PDE governing linear, adiabatic, nonraradial oscillations in a
spherical, differentially and slowly rotating non-magnetic star is derived.
This equation describes mainly low-frequency and high-degree $g$-modes,
convective $g$-modes, and rotational Rossby-like vorticity modes and their
mutual interaction for arbitrarily given radial and latitudinal gradients of
the rotation rate. Applying to this equation the `traditional approximation'
of geophysics results in a separation into radial- and angular-dependent
parts of the physical variables, each of which is described by an ODE. The
condition for the applicability of the traditional approximation is
discussed.
The angular parts of the eigenfunctions are described by Laplace's tidal
equation generalized here to take into account differential rotation. From
a qualitative analysis of Laplace's tidal equation the sufficient condition
for the formation of the dynamic shear latitudinal Kelvin-Helmholtz
instability (LKHI) is obtained. A small rotation gradient causes LKHI of
prograde waves (seen in the rotating frame), while strong gradients are
responsible for retrograde LKHI. The value of the latitudinal rotation
gradient has a lower limit, below which LKHI disappears.
The LKHI result is applied to real solar helioseismology rotation data.
It is shown that the $m=1$ mode ($m$ = azimuthal wave number) instability
can develop. This global instability takes place in the whole envelope of the
Sun, including the greatest part of the tachocline, in radial direction and
at almost all latitudes in horizontal direction.
The exact solutions of Laplace's equation for low frequencies and rigid
rotation are obtained. There exists only a retrograde wave spectrum in this
ideal case. The modes are subdivided into two branches: fast and slow modes.
The long fast waves carry energy opposite to the rotation direction,
while the shorter slow-mode group velocity is in the azimuthal plane
along the direction of rotation. The eigenfuncions are expressed by
Jacobi's polynomials which are polynomials of higher order than the
Legendre's for spherical harmonics.
The solar 22-year mode spectrum is calculated. It is shown that the slow
22-year modes are concentrated around the equator, while the fast modes are
around the poles. The band of latitude where the mode energy is concentrated 
is narrow, and the spatial place of these band depends on the wave numbers
($l, m$). %
}

\titlerunning{Eigenoscillations of the Sun: II.} 
\authorrunning{N.S. Dzhalilov \& J. Staude}
\maketitle


      \section{Introduction}


In a recent paper Dzhalilov et al. (2001; paper~1) investigated which
lowest-frequency eigenoscillations can occur in the real Sun, moreover,
which role they play in redistributing angular momentum and causing solar
activity. We found that such waves could only be differential rotation
Rossby-like vorticity modes. However, the general nonradial pulsation theory
adopted from stellar rotation has some difficulities. For slow rotation, when
the sphericity of the star is violated not seriously, the degeneracy of the
high-frequency spherical $p$-  and $g$-modes with respect to the azimuthal
number $m$ is abandoned by rotation (Unno et al. 1989). Independent of the
spherical modes non-rotating toroidal flows (called `trivial' modes with a
zero frequency) become quasi-toroidal with rotation (called $r$-modes with a
nonzero frequency; Ledoux 1951; Papaloizou \& Pringle 1978; Provost et al.
1981;  Smeyers et al. 1981; Wolff 1998). Although rotation abandons the
degeneracy of the modes, it also couples the modes with the same
azimuthal order, and this makes the problem more difficult. For the
high-frequency modes ($\en_{\sR}=\omega/2\Omega\ge 1$, where  $\omega$ and
$\Omega$ are the angular frequencies of oscillations and of stellar rotation,
respectively) this difficulty is resolved more or less successfully. For this
case the small perturbation rotation theory is applied, in  which the
eigenfunctions are represented by power series, the angular parts of which
are expressed by spherical harmonic functions $Y^m_l$ (Unno et al. 1989).
These power series are well truncated, unless $\en_{\sR}<1$, when the role of
Coriolis  force is increasing.

Namely the low-frequency instabilities are discovered in most
pulsating stars (Cox 1980; Unno et al. 1989). Rotation couples strongly
together the high-order $g$, the convective $g$, and the $r$-modes with
$\en_{\sR}<1$ and with the same $m$, but different $l$ (Lee \& Saio 1986).
Generally the matrix of the coupling coefficients to be determined is
singular (e.g. Townsend 1997). In all papers on the eigenvalue problem of
nonradially pulsating stars, there exists a `truncation problem' for the
serial eigenfunctions, the angular parts of which are represented by
spherical harmonics (e.g. Lee \& Saio 1997; Clement 1998).  

The governing partial differential equations (PDEs) of the eigenoscillations
of rotating stars are complicated from the point of view of the mathematical
treatment, even if the motions  are adiabatic. This difficulty arises because
in spherical geometry an eigenvalue problem with a singular boundary
condition has to be solved. These equations are simplified considerably
to neglect the tangential components of the angular velocity $\vec\Omega$ in
the low-frequency case $\en_{\sR}<1$ (this means that the motion caused by
the Coriolis force is primarily horizontal). This limitation widely used
in geophysical hydrodynamics (e.g. Eckart 1960) is called `traditional
approximation' and has been used first by Laplace (1778) to study tidal waves
(Lindzen \& Chapman 1969). Laplace's equation (or the traditional
approximation) for $\en_{\sR}<1$ is applicable to the stellar case too. The
main advantage of this approximation is that it decomposes the initial system
of equations into a pair of ordinary differential equations (ODEs) (e.g.
Lindzen \& Chapman 1969; Berthomieu et al. 1978; Bildsten et al. 1996; Lee \&
Saio 1997). The angular parts of the eigenfunctions are described by
Laplace's tidal equation. Solving this equation numerically by using a
relaxation method, Lee and Saio (1997) first avoided the representation of
the solutions by $Y_l^m(cos\theta)$ functions for the $\Omega=$const case,
and they had no problem with the truncation of the series. 

In the present work for the non-magnetic and non-convective cases we receive
one PDE in spherical geometry for the adiabatic pressure oscillations in the
differentially rotating star ($\Omega=\Omega(r, \theta)$) with arbitrary
spatial gradients of rotation (Sect.~2). This general equation is split into
the $\theta$- and $r$-component ODEs, if the traditional approximation is
applied (Sect.~3). The $\theta$-component equation is Laplace's tidal
equation generalized for the differentially rotating case. In Sect.~4 we
analyse more qualitatively this equation. We find the general condition
for the shear instability due to differential rotation in latitude.
We find that the smallest rotation gradient is responsible for the prograde
(seen in the rotating frame) vorticity wave instability, while a stronger
gradient causes the retrograde wave instability. For solar data (small
rotation gradients) the $m=1$ prograde mode instability is possible
(Sect.~4.4). The possible existence of such a global horizontal shear
instability on the Sun has been investigated by Watson (1981) and Gilman \&
Fox (1997), that of shear and other dynamic instabilities and of thermal-type
instabilities in stars as well by Knobloch \& Spruit (1982) and others.
Laplace's tidal equation for low frequencies in the rigid-rotation case is
investigated in detail in Sect.~5. It is shown that the eigenfunctions are
defined by Jacobi's polynomials which are of higher order than the
Legendre's.

\section{Basic equations}      

The fluid motion in a self-gravitating star, neglecting a magnetic field
and viscosity, may be described in an inertial frame by the hydrodynamic
equations. These equations in conventional definition are written as
\vspace{1mm}
\begin{eqnarray}
\frac{\partial \rho}{\partial t} + \nabla \cdot (\rho \vec{V})&=&0
,\label{1}\\
\rho \,\left(\frac{\partial}{\partial t} + \vec{V} \cdot \nabla
\right) \,\vec{V}&=& - \nabla p - \rho \nabla \Phi, \label{2} \\
\rho\,T\left(\pd{}{t}+\vec{V} \cdot \nabla \right)s&=& \rho
\,\varepsilon_{\sN} - \nabla \cdot  \vec{F}, \label{3en} \\
\nabla^2 \,\Phi&=& 4\pi G\,\rho. \label{3}                
\end{eqnarray}  

\subsection{Equilibrium state}	

We suppose that the equilibrium state (variables with zero indices) of the 
star is stationary and that its differential rotation is axially symmetric:
\be
\vec{V}_0(r,\theta) = \vec{\Omega}\times\vec{r} = 
\Omega \,r \,\sin\theta \,\vec{e}_{\phi},
\ee
where $\vec{\Omega}(r,\theta) = \Omega_r \,\vec{e}_r + \Omega_{\theta} \,
\vec{e}_{\theta}$, with the components $\Omega_r = \Omega\cos\theta, 
\ \ \Omega_{\theta} = - \Omega\sin\theta,$ and $\Omega_{\phi} \equiv 0 $, 
is the stellar angular velocity of rotation described in spherical
polar coordinates, $(r, \theta, \phi)$. Here $\vec{e}_{i}$ with
$i=r, \theta,$ or $\phi$ are the unit vectors. We will not include convective
motion and meridional flows into the initial steady state. In that case we
may obtain, in particular from the Eq.(\ref{2}) of motion, the hydrostatic
equilibrium relation 
\be -\frac{\nabla \,p_0}{\rho_0} = \nabla\left(\Phi_0 -
\frac{1}{2} \,|\vec{\Omega}\times \vec{r}|^2 \right) + \Omega \,r^2 \sin^2 
\theta \, \nabla \Omega = \vec{\tilde{g}} . \label{17} 
\ee
It follows that the effective gravity $\vec{\hbox{\~{g}}}$ cannot be a
potential field if differential rotation $\nabla\Omega\ne 0$ is present. This
is important for rapidly rotating stars where the configuration is deformed
by the centrifugal force as well as by differential rotation. For slowly
rotating stars (the Sun as well) we may assume that the initial state is only
marginally disturbed by rotation and $\vec{\tilde{g}}  \approx
\vec{g}=\nabla\Phi_0$ can be applied. That is, non-sphericity is not
essential for the generation of waves (Unno et al. 1989).

\subsection{Equations of oscillation} 

Small amplitude deviations from the basic state of the star may be
investigated by linearizing Eqs.~(\ref{1}--\ref{3}). For Eulerian
perturbations (variables with a prime) the equation of motion becomes 
(Unno et al. 1989) 
\begin{eqnarray}
\vec{e}_i\frac{\partial}{\partial t^{\prime}} \,V_i + 2\, \vec{\Omega} \times 
\vec{V} + \vec{e}_{\phi}\, r \,\sin \theta (\vec{V} \cdot \nabla
\Omega)=  \nonumber\\
= \frac{\nabla p_0}{\rho_o} \,
\frac{\rho^{\prime}}{\rho_0} - \frac{1}{\rho_0} \,\nabla p^{\prime} 
-  \nabla \Phi^{\prime}. \label{20}
\end{eqnarray}     
Here the operator
\begin{equation}
\frac{\partial}{\partial t^{\prime}} = \frac{\partial}{\partial t} + \Omega
\,\frac{\partial}{\partial \phi} \label{21}
\end{equation}
represents the temporal derivative referring to a local frame rotating with
an angular velocity $\Omega=\Omega(r,\theta)$. For low-frequency waves Saio
(1982) has shown numerically in detail, that the Cowling approximation is
good enough in most cases. Thus we will neglect perturbation of the
gravitational potential in Eq.~(\ref{20}), $\Phi^{\prime}=0$. We are
interested in very slow motions such that $v_{\rm ph} \ll c_{\rm s}$, where
$v_{\rm ph}$ is the phase velocity of the waves and $c_{\rm s}$ is the sound
speed. Then the incompressible fluid motion limit,  $c_{\rm s}^2\to\infty$
(it is within the adiabatic approximation) may be applied, and instead
of the Eq.~(\ref{1}) of mass conservation we use  
\be
\nabla\cdot\vec{V}\!=\!\frac{1}{r^2}\pd{(r^2V_r)}{r}\!+\!
\frac{1}{r\sin\theta}\pd{(V_{\theta}\sin\theta)}{\theta}\!  +\!
\frac{1}{r\sin\theta}\pd{V_{\phi}}{\phi}\!=\!0 . \label{23} 
\ee 
We have shown in Paper~1, that nonadiabatic effects are of great importance
for the dynamics of low-frequency rotation modes. However, here we shall
restrict ourselves to the adiabatic case only because the mathematical
treatment of wave equations in spherical geometry is rather difficult.
For adiabatic waves we receive from Eq.~(\ref{3en}) 
\be
\frac{\partial \rho^{\prime}}{\partial t^{\prime}} - \rho_0 \,\frac{N^2}{g}
\,V_r = \frac{1}{c^2_s} \,\frac{\partial p^{\prime}}{\partial t^{\prime}} ,
\label{27} 
\ee      
where the squared Brunt-V\"ais\"al\"a frequency  
\begin{eqnarray}
N^2 = g \left( \frac{1}{\Gamma_1} \,\frac{1}{p_0} \,\frac{d p_0}{d r} - 
\frac{1}{\rho_0} \,\frac{d \rho_0}{d r} \right)
\end{eqnarray}         
is written for the slow rotation case where $\tilde g\approx g$. For rapid
rotation $N^2=N^2(r,\theta)$ and $g$ should be changed here to $\tilde g$. In
the incompressible limit ($c_s^2\to\infty$) Eq.~(\ref{27}) reads
\be \frac{\partial \rho^{\prime}}{\partial t^{\prime}} = \rho_0 \,
\frac{N^2}{g} \,V_r . \label{28}
\ee
Thus we have a complete set of Eqs.~(\ref{20}, \ref{23}, \ref{28}) 
to describe  adiabatic, low-frequency, non-radial oscillations in a
differentially rotating star. For our axisymmetric stationary initial
state we may represent all the perturbed variables $\vec V$, $p'$, and
$\rho'$ in the inertial frame as 
\be
\vec{V}(r, \theta, \phi; t) \Longrightarrow \vec{V} \,(r, \theta) \, 
e^{i(m \phi-\omega_0 t)} . \label{29}  
\ee
 Considering $\pdtext{}{t'}=-i\omega$ and $\pdtext{}{\phi}=im$ we find
from Eq.~(\ref{21}) the relation between the frequencies in the inertial and
rotating frames 
\be
\omega = \omega_0 - m\,\Omega\,(r, \theta) , \label{30}
\ee
from where we get $\nabla\omega=-m\nabla\Omega$. If we separate
the variable part of the rotation frequency,
$\Omega(r,\theta)=\Omega_{\odot}+ \tilde\Omega(r,\theta)$, then
$\omega=\omega_0 - m\Omega_{\odot} - m\tilde\Omega = \omega_{\odot} -
m\tilde\Omega =\omega_{\odot} -  k_{\phi}v_{\rm ph}$ ($k_{\phi}$ and $v_{\rm
ph}$ are the local azimuthal wave number and the phase velocity,
respectively). We will study the case $\tilde\Omega \ll
\Omega_{\odot}$. From now $\Omega\approx \Omega_{\odot}$ and $\nabla\Omega
= \nabla\tilde\Omega\ne 0$ will be used. Low frequencies seen in the
rotating frame mean that we are close to the resonant frequencies in the
inertial frame ($\omega_{\odot}\approx  k_{\phi}v_{\rm ph}$).

Now excluding $\rho'$ from Eq.~(\ref{20}) by using Eq.~(\ref{28}), taking the
projection of this equation onto the rotation axis $\vec\Omega$ and two
tangential components as well, and adding Eq.~(\ref{23}) we get 
\begin{eqnarray}
\left( 1 - \frac{N^2}{\omega^2} \right) V_r \cos \theta - j V_{\theta} \sin
\theta = \nonumber\\
=\frac{1}{\rho_0 i \omega} \left( \cos \theta \frac{\partial}{\partial
r} - j \frac{\sin \theta}{r} \,\frac{\partial}{\partial \theta} \right)
p^{\prime}, \label{34} \\
V_{\theta} + \frac{2 \Omega \cos \theta}{i \omega} V_{\phi} = 
\frac{1}{r \rho_0 i \omega} \,\frac{\partial p^{\prime}}{\partial \theta},
\label{35} \\ 
- i \omega V_{\phi} + j \sin \theta \left( 2 \Omega + r \frac{\partial
\Omega}{\partial r} \right) V_r + \nonumber\\
+ \left( 2 \Omega \cos \theta + \sin \theta 
\frac{\partial \Omega}{\partial \theta} \right) V_{\theta} = - \frac{i m}{r
\rho_0 \sin \theta}  p^{\prime}, \label{36} \\   
i m V_{\phi} + \frac{\sin \theta}{r} \,\frac{\partial}{\partial r} (r^2 V_r)
+ \frac{\partial}{\partial \theta} (\sin \theta V_{\theta}) = 0. \label{37}  
\end{eqnarray}               
Here we have introduced the special parameter $j$ to switch to the
traditional approximation ($\Omega_{\theta}=0$, $\Omega_r\ne 0$).
In the general case $j\equiv 1$, and for switching to the traditional
approximation we put $j=0$. Further we will obtain one additional equation
for $p'$.

Let $\mu=\sin\theta$ be a new independent variable and 
\begin{eqnarray}
\en_{\sR} &=&  \frac{\omega}{2 \Omega} , \ \beta_r =
\frac{r}{2\Omega} \frac{\partial \Omega}{\partial r} , \ \beta_{\mu} =
\frac{\mu}{2\Omega} \frac{\partial \Omega}{\partial \mu}, \  \nonumber\\
\alpha &=& \varepsilon^2_{\sR} - (1 - \mu^2)(1+ \beta_{\mu}), \\
a_1 &=& \frac{1 + \beta_r}{\varepsilon_{\sR}}, \  \ a_2 = \frac{1 +
\beta_{\mu}}{\varepsilon_{\sR}}, \  \ a_3 =
\frac{\alpha}{\varepsilon^2_{\sR}},  \nonumber\\ 
a_4 &=& \mu^2 \frac{1 + \beta_r}{\alpha} ,  
a_5 = \frac{m}{\varepsilon_{\sR}} (1 + \beta_{\mu}) - 1 . \nonumber
\end{eqnarray}  
$\en_{\sR}$ is the Rossby number; we are interested in $\en_{\sR}\le 1$.
Using 
\begin{equation}
i V_{\phi} = j a_1 \mu V_r + a_2 V_{\theta} \cos \theta - \frac{m}{\rho_0 i
\omega} \, \frac{p^{\prime}}{r \mu} \label{39} 
\end{equation}   
we get three equations from the Eqs.~(\ref{34}--\ref{37}) for $V_r$,
$V_{\theta}$, and $p'$. From those $V_{\theta}$ is excluded by
\begin{equation}
\frac{V_{\theta}}{\cos\theta} = j a_4 \tilde{V}_r - \frac{1}{a_3} 
\left( a_5 - \mu \frac{\partial}{\partial \mu} \right) \tilde{P} , \label{47}
\end{equation}
where
\begin{equation}
p^{\prime} = \mu r \rho_0 i \omega \tilde{P} \ , \,\, V_r = \mu \tilde{V}_r ,
\label{43} 
\end{equation}          
and we get two equations for $\tilde{V}_r$ and $\tilde{P}$:
\begin{eqnarray}
&& \left( 1 - \frac{N^2}{\omega^2} - j a_4 \right) \tilde{V}_r = \nonumber\\
&&=\left[ b_1 + r\frac{\partial}{\partial r} - 
j a_6 - j \left( 1 - \frac{1}{a_3} \right) \mu 
\frac{\partial}{\partial \mu} \right] \tilde{P}, \label{48}\\
&&\left[ 2 + r \frac{\partial}{\partial r} + j \left( a_8 + a_9 \mu
\frac{\partial}{\partial \mu} \right) \right] \tilde{V}_r = \check{A}(\mu)
\tilde{P}, \label{49}\\
&&\check{A} (\mu)\!=\!\frac{1}{\mu^2} \left[ a_7\!+\!(1\!-\!\mu^2) \mu
\frac{\partial}{\partial \mu} \right]\! \frac{1}{a_3}\!\left( a_5\!-\!\mu
\frac{\partial}{\partial \mu} \right)\!+\!\frac{m^2}{\mu^2}. \nonumber
\end{eqnarray}
\vspace{2mm}
Here the dimensionless coefficients are defined as
\vspace{1mm}
\begin{eqnarray}
a_6 &=& \frac{1}{\alpha} \left[ m \varepsilon_{\sR}\!+\! (1\!-\!\mu^2)(1\!+\!
\beta_{\mu}) (\frac{m}{\varepsilon_{\sR}} \beta_{\mu}\!-\!1) \right] ,
\nonumber\\
a_7 &=& \frac{m}{\varepsilon_{\sR}} ( 1 + \beta_{\mu}) (1 - \mu^2) + 1 - 2
\mu^2 \ , \nonumber\\
a_8 &=& \frac{a^*_8}{\alpha^2} \ ; \ a_9 = \frac{1}{\alpha} (1 - \mu^2) 
(1 + \beta_r),   \\
a_8^*  &=& (1+\beta_r)\left[\alpha (3-4\mu^2 + m\varepsilon_{\sR})\right. -  
\nonumber\\
&& \left.
-(1-\mu^2)\check{\delta}_{\mu}\alpha\right]-2\alpha(1-\mu^2)\beta_r \beta_{\mu} , \nonumber\\ 
b_1 &=& 1 +  \kap_{\rho} - \frac{m}{\varepsilon_{\sR}} \beta_r,\,\,
\kap_{\rho} = \frac{r}{\rho_0} \,\frac{d \rho_0}{d r}. \nonumber  
\end{eqnarray}
Deriving these equations the second derivatives $\Omega''(\theta)$ and
$\Omega''(r)$ have been omitted as very small quantities. Eq.~(\ref{48})
allows to obtain one equation for $p'$:
\be
\Big[\psi_1 \check{\delta}^2_r + \psi_2 \check{\delta}^2_{\mu} + \psi_3 
\check{\delta}_r + 
\psi_4 \check{\delta}_{\mu} + \psi_5 \check{\delta}_r \check{\delta}_{\mu} 
+ \psi_6 \Big] \tilde{P} = 0 . \label{50}
\ee
This is the main singular PDE for nonradial rotation-gravity waves in a
differentially rotating star. The coefficients  $\psi_{1-6}$ of
Eq.~(\ref{50}) are rather complicated, we  present them in Appendix A. The
operators are defined as 
\be
\check{\delta}_r =r\pd{}{r} , \  \check{\delta}_{\mu}= \mu\pd{}{\mu} , \ 
{\rm and} \  \check{\delta}^2_r =\check{\delta}_r\check{\delta}_r . 
\ee
First we will study this equation in different simplified approximations.
The most popular case is the traditional appraximation.

\section{The `traditional approximation': $j \equiv 0$}      

Strictly speaking, the condition $j=0$ in Eq.~(\ref{50}) is not applicable in
two points: (1) at $\omega^2=N^2(r)$, that is in the turning points in radial
direction; (2) at $\alpha=0$ or $\en^2_{\sR}\approx\cos^2\theta$, that is in
the latitudinal turning point. The last one is more important, because the
traditional approximation filters out such important and interesting
phenomena as the trapping of Rossby-like waves around the equator. In
geophysics this phenomenon is investigated separately (Pedlosky 1982; Gill
1982). The applicability of the traditional approximation for rigid
rotation has been checked by numerical modelling as well as by experimental
verification. For minor differential rotation of the star (small $\beta_r$
and $\beta_{\mu}$) these examinations are also valid. Thus for $j=0$ 
Eq.~(\ref{50}) becomes:
\begin{eqnarray}
\frac{\varepsilon_{\sR}}{\alpha^2 \mu^2} \left[ \alpha \varepsilon_{\sR} (1 -
\mu^2) \mu^2 \frac{\partial^2}{\partial \mu^2} + q_3 \mu
\frac{\partial}{\partial \mu} + q_4 \right] p^{\prime}=  \nonumber\\
= - \frac{1}{\psi} \left( r^2\frac{\partial^2}{\partial r^2} + b_4 r 
\frac{\partial}{\partial r} + b_5 \right) p^{\prime} \label{54} 
\end{eqnarray}
with the parameters
\begin{eqnarray}
q_1 &=& 2 - \mu^2 - \varepsilon^2_{\sR} \frac{2 - 3\mu^2}{1 - \mu^2} +
\nonumber\\  
&+& \beta_{\mu} \left[ 2 \mu^2 - 1 - 2 (1 - \mu^2)(1 + \beta_{\mu})
\left( m/\en_{\sR} +1 \right) \right],    \nonumber\\
q_2 &=& h_0 + h_1 \beta_{\mu} + h_2 \beta^2_{\mu}, \nonumber\\
h_0 &=& \varepsilon_{\sR} (m^2 - 1)(\varepsilon^2_{\sR} + \mu^2 - 1) -
\nonumber\\
 &-& \mu^2 [m (1 - \mu^2) + \varepsilon^2_{\sR} (m - 2 \varepsilon_{\sR})],
 \nonumber\\
h_1 &=& (1 - \mu^2) \left[ - 2 m \mu^2 + \varepsilon_{\sR} (m^2 - 4 + 
5 \mu^2) \right. + \nonumber\\
 &+& \left. 4 m \varepsilon^2_{\sR} \right] - m \varepsilon^2_{\sR},
 \nonumber\\
h_2 &=& (1 - \mu^2)^2 (3m - 2 \varepsilon_{\sR} + 2 m \beta_{\mu})-\nonumber\\
&-&(1 - \mu^2) m (1 - 2 m \varepsilon_{\sR}), \nonumber\\
b_2 &=& 3 + \kap_{\rho} - \beta_r \frac{m}{\varepsilon_{\sR}} -
\frac{\check{\delta}_r \psi}{\psi}, \nonumber\\
b_3 &=& 2 + 3\kap_{\rho} - \beta_r \frac{m}{\varepsilon_{\sR}} \left(
3 + \beta_r \frac{m}{\varepsilon_{\sR}} \right) - \nonumber\\
&-&\left( 1 + \kap_{\rho} - \beta_r
\frac{m}{\varepsilon_{\sR}} \right) \frac{\check{\delta}_r \psi}{\psi},
\nonumber\\
\psi &=& 1 - \frac{N^2}{\omega^2} \ , \ \check{\delta}_r \psi = - 2
\frac{N^2}{\omega^2} \,\frac{m}{\varepsilon_{\sR}} \beta_r -
\frac{r}{\omega^2} \, \frac{d N^2}{d r} 
\end{eqnarray}
\begin{eqnarray}
q_3 & = & \varepsilon_{\sR} (1 - \mu^2) [\alpha (1-2 a_0) - q_1] , \ a_0 = 1
- \beta_{\mu} \frac{m}{\varepsilon_{\sR}} , \nonumber\\
q_4 & = & \varepsilon_{\sR}(1-\mu^2) [\alpha(a_0^2-\check{\delta}_{\mu} a_0)
+a_0 q_1]-q_2 , \nonumber\\
\check{\delta}_{\mu}a_0 & = &-\beta_{\mu}
\frac{m}{\varepsilon_{\sR}}\left(1+\beta_{\mu}
\frac{m}{\varepsilon_{\sR}}\right) , \nonumber\\
b_4 &=& 1\!-\!2b_1\!+\!b_2 , \ \check{\delta}_r b_1
\!=\! \kap_{\rho}\!-\!\beta_r \frac{m}{\varepsilon_{\sR}}\left(1+\beta_r
\frac{m}{\varepsilon_{\sR}}\right) , \nonumber\\
b_5 &=& b_1^2-\check{\delta}_r b_1 - b_1 b_2 + b_3 . \nonumber
\end{eqnarray}
Remembering that $\tilde\Omega(r,\theta) \ll \Omega_{\odot}$, the
left-hand side of Eq.~(\ref{54}) is a function of $\mu=\sin\theta$, while the
right-hand side is a funcion of $r$ only. In that way we may separate
the variables
\begin{equation}
   p'(r, \mu) = \Theta(\mu)Q(r). \label{55}
\end{equation}
Now putting Eq.~(\ref{55}) into (\ref{54}) we receive the `$\theta$'- and
`$r$'-equations:
\begin{equation}
\left[\alpha\varepsilon_{\sR}(1\!-\!\mu^2)\mu^2\frac{d^2}{d\mu^2}\!+\!
q_3\mu\frac{d}{d\mu}\!\!+\!\!q_4+
\Lambda\frac{\alpha^2\mu^2}{\varepsilon_{\sR}} \right]\Theta(\mu)\!=\! 0 , 
\label{56}
\end{equation}
\begin{equation}
\left(r^2\frac{d^2}{dr^2}+b_4 r\frac{d}{dr}+b_5 -\Lambda\psi\right)Q(r)=0.
\label{57}
\end{equation}
These two equations are connected to each other by two common spectral
parameters: $\omega$ -- the oscillation frequency and $\Lambda$ --
the separation parameter. Both must be searched for as a solution of
the boundary value problem. So far the logarithmic gradients of the
rotation rate are arbitrary functions, $\beta_{\mu}=\beta_{\mu}(\mu)$,
$\beta_r=\beta_r(r)$.

\section{Generalized Laplace's tidal equation}   

Eq.~(\ref{56}) is the generalized Laplace equation if differential rotation is
present, $\beta_{\mu} \neq 0$. For rigid rotation, $\beta_{\mu} = 0$,
Eq.~(\ref{56}) becomes the standard Laplace equation:
\vspace{1mm}
\begin{eqnarray}
\left[\frac{1-\mu^2_{\ast}}{\varepsilon^2_{\sR}-\mu^2_{\ast}}
\frac{d^2}{d\mu^2_{\ast}} + \frac{1-\varepsilon^2_{\sR}}{(\varepsilon^2_{\sR}
- \mu^2_{\ast})^2}2\mu_{\ast}\frac{d}{d\mu_{\ast}}\right. &-& \nonumber\\
-\left.\frac{1}{\varepsilon^2_{\sR} -
\mu^2_{\ast}}\left(\frac{m^2}{1-\mu^2_{\ast}} -
\frac{m}{\varepsilon_{\sR}}\frac{\varepsilon^2_{\sR}+\mu^2_{\ast}}
{\varepsilon^2_{\sR}-\mu^2_{\ast}}\right) +
\frac{\Lambda}{\varepsilon^2_{\sR}} \right]\Theta &=& 0, \label{58}  
\end{eqnarray}
\vspace{1mm}

\noindent
where $\mu_{\ast}=\cos\theta$. A peculiarity of this equation is the presence
of three singular points: at the pole, at the equator, and between both if
$\en^2_{\sR}\leq 1$ (our case). Therefore it is hard to solve such an
equation analytically or numerically to find the eigenvalues. Since the
time of Laplace in geophysics investigations were focused on almost
two-dimensional (horizontal) motions in strongly stratified fluids with 
$V_r\approx 0$. In this situation the $r$-component Eq.~(\ref{57}) does
not appear, and one is looking for the eigenvalues $\Lambda$ in Laplace's
Eq.~(\ref{58}) for a given $\en_{\sR}$ (it is expressed through the
thickness of the fluid layer, e.g., in a shallow water-wave system).
For the special cases and for the general case too, when the eigenfunctions
are expressed through the Hough functions (essencially these are the same
infinite series of $Y_l^m$ harmonics) references could be found in a paper by
Lindzen \& Chapman (1969). 

In astrophysics the $r$-component Eq.~(\ref{57}) appears, and two equations
must be solved together to find both spectral parameters. In the
rigid-rotation case Lee and Saio (1997) looked for $\Lambda$ numerically in
an approach similar to that in geophysics, fixing $\en_{\sR}$ in
Eq.~(\ref{58}). We offer here another approach, where we will find $\Lambda$
from the $r$-equation for a given $\en_{\sR}\leq 1$.

It is convenient to introduce into Eq.~(\ref{56}) the new variable  
$x=\mu^2=\sin^2\theta$:
\begin{equation}
4(1\!+\!\beta_{\mu})(1\!-\!x)(x\!-\!a)x^2\frac{d^2\Theta}{dx^2}\!-\!A_1
2x\frac{d\Theta}{dx}\!+\!A_2\Theta\!=\!0 , \label{59}
\end{equation}
where
\begin{eqnarray}
A_1\!&=&\!(1\!-\!x)[2\!-\!x\!-\!3\varepsilon^2_{\sR}\!+\! 
      \beta_{\mu}(4x\!-\!3\!-\!2m\en_{\sR})]\!-\!  \nonumber \\
\!&-&\!2\beta_{\mu}^2 (1\!-\!x)^2\!+\!\varepsilon^2_{\sR}, \label{60} \\
A_2\!&=&\!(1\!-\!x)[m^2
(1\!-\!\beta_{\mu})\!+\!\frac{m}{\varepsilon_{\sR}}x\!
+\!\beta_{\mu}\frac{m}{\varepsilon_{\sR}}(4x\!-\!3\!-\!2\en^2_{\sR})]\!-\!
\nonumber
\\
\!&-&\!m\varepsilon_{\sR}(1\!-\!x)\!\!+\!
m\varepsilon_{\sR}(1\!-\!m\varepsilon_{\sR})\!+\!\frac{\Lambda}{\en^2_{\sR}}
x(x\!-\!a)^2(1\!+\!\beta_{\mu})^2, \nonumber
\\
a\!&=&\!1-\frac{\varepsilon^2_{\sR}}{1+\beta_{\mu}} . \label{65} 
\end{eqnarray}
This equation determines the tangential structure of the eigenfunctions,
while Eq.~(\ref{57}) is responsible for the radial behavior. A detailed
investigation of Eq.~(\ref{57}) is not included in this work. A similar
equation for a realistically stratified model of the Sun has been
investigated by Oraevsky \& Dzhalilov (1997). We remind that the radial
structure of the eigenfunctions depends on the sign of $\Lambda$ (either
radiative or convective modes). Now we can already estimate an approximate
value of $\Lambda$:  
\be
\frac{\Lambda}{\varepsilon^2_{\sR}} \sim \frac{n^2\pi^2}
{N^2_m\en^2_{\sR}/\omega^2}  =
\frac{n^2\pi^2}{N^2_m/4\Omega^2} \sim n^2\times 10^{-6}, \label{61}
\ee
where $n$ is the radial harmonic number, $N_m$ is mean value of the
Brunt-V\"ais\"al\"a frequency in the radiative interior or in the convection
zone, and $N_m/2\Omega$ is the Prandtl number. An estimate of Eq.~(\ref{61})
is done for a solar model, but we think similar values of the Prandtl number
are valid for most other stars too. Thus we can omit from $A_2$ in
Eq.~(\ref{60}) the last term, if the radial number $n$ is not too large.  

\subsection{Fluid velocities}   

From Eqs.~(\ref{39}, \ref{47}, \ref{48}) we can derive in the traditional
approximation the following formulae for the components of the
fluid velocity:
\begin{equation}
V_r = \frac{1}{\rho_0 i\omega\psi} \pd{p'}{r}, \label{62} 
\end{equation}
\begin{eqnarray}
V_{\theta}\!= \frac{\pm 1}{r \rho_0 i\omega}
\frac{\varepsilon^2_{\sR}}{1\!+\!\beta_{\mu}} \frac{1\!-\!x}{a\!-\!x}
\frac{1}{\sqrt{x(1\!-\!x)}}
\left(\frac{m}{\varepsilon_{\sR}}\!-\!2x\pd{}{x}\right)\!p', \label{63}
\end{eqnarray}
\begin{equation}
V_{\phi} = \frac{1}{r \rho_0 \omega} \frac{1}{\sqrt{x}}
\left[\frac{1-x}{a-x}\left(2\varepsilon_{\sR} x\pd{}{x} - m\right) +
m\right]p'. \label{64}
\end{equation}
\vspace{1mm}
Here the different signs $\pm$ of $V_{\theta}$ correspond to the
northern and southern hemispheres, so that $\cos\theta=\pm\sqrt{1-x}$. Our
further aim is to find such solutions for $p'=\Theta(x)Q(r)$ that all the
components of the velocity remain limited at the pole ($x=0$), at the equator
($x=1$), and in both turning points, where $x=a$ and $\psi(r)=0$. 

\subsection{Heun's equation}   

Now we will impose a restriction to $2\beta_{\mu} =
\pdtext{(\ln\Omega)}{(\ln\mu)} \approx$ const, the logarithmic latitudinal
gradient of the rotation frequency. We might take the linear dependence
$\beta\sim x$, but for such a profile the structure of the solutions is not
changed qualitatively.  Let us introduce the new dependent variable
\begin{equation}
\Theta = x^{\sigma}Y(x) , \label{66}
\end{equation}
where
\begin{eqnarray}
2\sigma &=& \beta_{\mu}S_1 \pm\sqrt{\beta^2_{\mu}S^2_1-S_2} = 2\sigma_{1, 2}
, \label{67} \\ 
S_1 &=& \frac{5+2m\varepsilon_{\sR} + 2\beta_{\mu}}{2(1+\beta_{\mu}
- \varepsilon^2_{\sR})} ,\nonumber\\
S_2 &=& \frac{\beta_{\mu}\frac{m}{\varepsilon_{\sR}}(3+2\varepsilon^2_{\sR})
- m^2(1-\beta_{\mu}-\varepsilon^2_{\sR})} {1+\beta_{\mu}-\varepsilon^2_{\sR}}
. \nonumber
\end{eqnarray}
Then for $Y(x)$ we get a new equation from Eq.~(\ref{59})
\begin{eqnarray}
x(1-x)(x-a)Y''\!&+&\!\frac{1}{2}\left[4\sigma(1-x)(x-a)\!-
\!\frac{A_1}{1+\beta_{\mu}}\right]Y'\!- \nonumber\\
&-&[(x-1)\nu_0 + \en_{\sR}\nu_1] Y = 0 , \label{68}
\end{eqnarray}
with
\begin{eqnarray}
\nu_0 &=& \frac{1+4\beta_{\mu}}{4(1+\beta_{\mu})}
\left(\frac{m}{\varepsilon_{\sR}} - 2\sigma\right)+
\sigma\beta_{\mu}(S_1\!-\!1) \!-\!\frac{S_2}{4} , \nonumber\\
\nu_1 &=& \frac{m(m\varepsilon_{\sR} -1) +
2\sigma\varepsilon_{\sR}}{4(1+\beta_{\mu})} . \label{68'}
\end{eqnarray}
\vspace{1mm}
\noindent
Eq.(\ref{68}) is Heun's equation (Heun 1889) in standard form
\begin{eqnarray}
x(x-1)(x-a)Y''+ \big[\gamma(x-1)(x-a) + \delta x(x-a) + \nonumber\\
\;\; + \varepsilon x(x-1)\big]Y' +
\tilde{\alpha}\tilde{\beta}(x-h)Y = 0 . \ \ \ \
\label{69}
\end{eqnarray}
\vspace{1mm}
The Riemannian scheme for this equation is
$$
p\left\{ \begin{array}{ccccc}
0 & 1 & a & \infty & \\ 
0 & 0 & 0 & \tilde{\alpha} & x \\
1\!-\!\gamma & 1\!-\!\delta & 1\!-\!\varepsilon & \tilde{\beta} &
\end{array}
\right\} , $$
where the exponents are connected by Riemann's relation $$ \tilde{\alpha}
+ \tilde{\beta} -\gamma - \delta - \varepsilon + 1 = 0 .$$
In the Riemann scheme the first row defines the singular points of Heun's
equation, while the corresponding exponents are placed in the second and
third rows. These exponents are
\begin{eqnarray}
1-\gamma &=& \beta_{\mu}S_1-2\sigma , \ \  1-\delta = \frac{1}{2} ,\\
1-\varepsilon &=& 2+\beta_{\mu}\left(1-S_1 +
\frac{3/2}{1+\beta_{\mu}}\right) , \nonumber\\
2\tilde{\alpha} &=& S\!+\!q , \  
2\tilde{\beta}\!=\! S\!-\!q ,\,\,\, q = \sqrt{S^2\!-\!4\nu_0} , \label{71} \\
h &=& 1-\varepsilon_{\sR}\frac{\nu_1}{\nu_0} , \ \  S\!=\!
2\sigma\!-\!2\!-\!\beta_{\mu}\!+\!\frac{3/2}{1\!+\!\beta_{\mu}} . \nonumber
\end{eqnarray}
Note that the second exponent at $x=a$ is $(1-\en)\to 2$ if $\en_{\sR}\to 0$
or if $\beta_{\mu}\to 0$. If $\beta_{\mu}\to\infty$ then $(1-\en)\to 7/2$.
That means, the second independent solution of Eq.~(\ref{69}) with the
exponent $(1-\en)$ is regular at the singular point $x=a$ for all variables,
which follows from Eqs.~(\ref{62}--\ref{64}). The exponent $(1-\delta) =
1/2$ also provides limited $V_\theta$ and $V_\phi$  at the equator ($x=1$).
The singularity $x=\infty$ in the Riemann scheme does not occur in our
task. The situation is more complicated around the pole $x=0$ with the
exponent $(1-\gamma)$. Let us consider this point in detail.

\subsection{Condition for the latitudinal Kelvin-Helmholtz instability}   

If we put the solutions with the exponents 0 and $(1-\gamma)$ into
Eq.~(\ref{66}) we get $p'\sim\Theta\sim x^{\sigma_{1,2}}$. Then
$V_{\theta,\phi}\sim x^{(2\sigma-1)/2}$ means that for the regularity of the
solutions at the pole the condition $Re(2\sigma)\ge 1$ must be obeyed. On the
other hand, an instability is possible when the eigenfrequencies are
complex, that means complex $\sigma$. For the latter it follows from
Eq.~(\ref{67}), that the necessary condition is $S_2>0$. It is clear that
the axially-symmetric mode with $m=0$ is excluded. For lower values of the
rotation gradient $|\beta_{\mu}|<1$ the necessary condition $S_2>0$ demands
for the prograde waves ($m\en_{\sR}>0$) the condition  $\beta_{\mu}>0$, which
is more realistic for stellar situations (equatorward spinning up at the
surface with radius $r$). Rayleigh's necessary condition for
instability (Rayleigh 1880; Watson 1981) says that the function
$Rl=\partial^2(\Omega\sin^2\theta)/\partial\cos^2\theta$ (gradient of
vorticity) must change its sign in the flow. Rewriting this function in our
definitions we get that
\be
Rl=\frac{2\Omega}{\mu^2}[3\beta_{\mu}-\mu^2(1+4\beta_{\mu})]  \nonumber
\ee
may change its sign if $\beta_{\mu}>0$. There are instability
possibilites for negative $\beta_{\mu}$ which are not considered in this
work. However, all formulas are valid for this case too.

The sufficient condition for instability is obtained from Eq.~(\ref{67})
and reads $\beta_{\mu}^2S_1^2 < S_2 $. The regularity condition at the pole
$\beta_{\mu}S_1 \ge 1$ can be rewritten as
\be
\frac{\varepsilon_{\sR}}{m} \le \frac{\beta_{\mu}\varepsilon_{\sR}^2}
{1-\varepsilon_{\sR}^2-\beta_{\mu}(\beta_{\mu}+3/2)} = \chi_3 . \label{72}
\ee
By this condition the phase space $\{\en_{\sR}/m, \en^2_{\sR}\}$ is
divided into three parts, depending on the values of $\beta_{\mu}$. For
$0\le\beta_{\mu}<1/2$ we have the following situation: if $\en_{\sR}^2 <
1-\beta_{\mu}(\beta_{\mu}+3/2)$  the condition Eq.~(\ref{72}) is fulfilled
for prograde waves $\en_{\sR}/m\ge 0 $ (region {\bf I}); in the
opposite case when $\en_{\sR}^2 > 1-\beta_{\mu}(\beta_{\mu}+3/2)$ the
condition (\ref{72}) is fulfilled for retrograde waves with
$\en_{\sR}/m < -1/2 $ (region {\bf III}); for $\en_{\sR}^2 =
1-\beta_{\mu}(\beta_{\mu}+3/2)$ these regions are separated by the
asymptote $\chi_3=\infty$.

\begin{figure}
\vspace{149mm}
\includegraphics{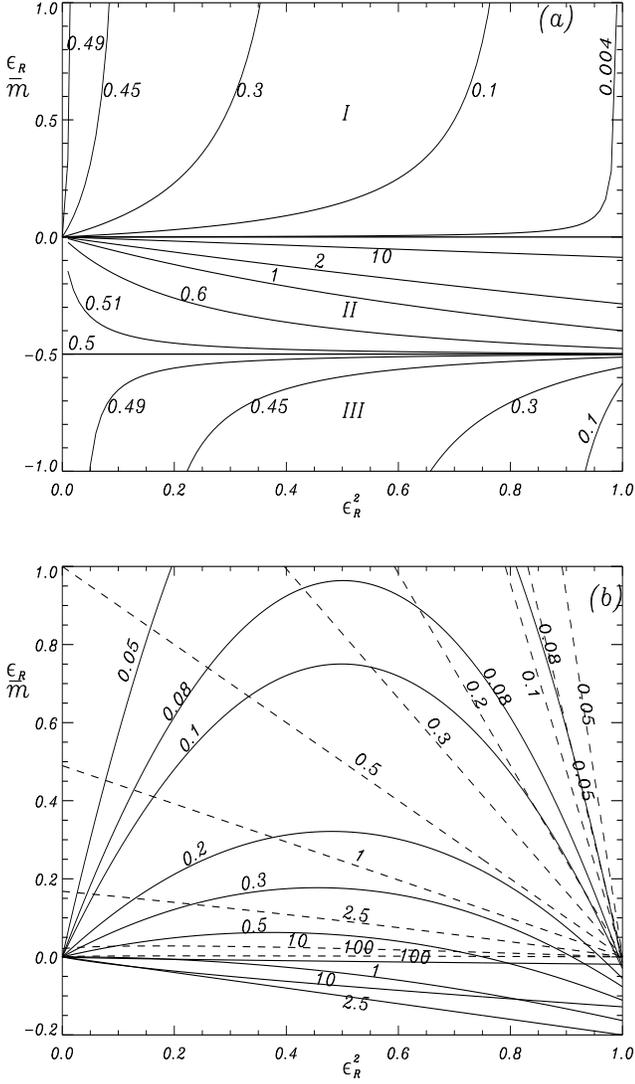}
\caption{The domains of validity of the solution regularity and of
the instability conditions in the phase space $\{\en_{\sR}/m,
\en^2_{\sR}\}$ for given values of the rotation gradient $\beta_{\mu}$.
(a) shows the values $\en_{\sR}/m = \chi_3$, Eq.~(\ref{72}). In
the area $\en_{\sR}/m \le \chi_3$ the solutions are limited at the
pole. (b) shows the behavior of $\chi_1$ (dashes) and $\chi_2$ (solid).
Between the solid and dashed lines with the same labels (values of
$\beta_{\mu}$) LKHI is possible.}
\end{figure}

For $\beta_{\mu} > 1/2$ (strong gradients) the condition Eq.~(\ref{72}) is
met only for retrograde waves in the range $-1/2 < \en_{\sR}/m \le 0$
(region {\bf II}). For $\beta_{\mu} = 1/2$ we get for any $\en_{\sR}$ that
$\chi_3=-1/2$. This is the line between the regions {\bf II} and {\bf III}.
All three regions are shown in Fig.~1a. It is seen that the regularity
condition is working for $|m|\ge 1$ and $|\en_{\sR}|\le 1$ if
$\beta_{\mu}\ne 0$. For very small $\beta_{\mu}$ only modes with large $m$
are possible. The smallest $m \gtrsim 1$ modes may appear in the limit
$\en_{\sR} \approx 1$. These conclusions are correct only if the
instability occurs.

Now let us consider the second condition, the complex frequency condition
$\beta^2_{\mu}S^2_1 < S_2$. This inequality may be rewritten as
\begin{eqnarray}
   \chi_2 &<&  \frac{\varepsilon_{\sR}}{m} < \chi_1 , \label{73}\\
   \chi_{1, 2}&=& \frac{2}{\beta_{\mu}(5+2\beta_{\mu})^2} \left[b_{\ast} \pm
   \sqrt{b_{\ast}^2-a_{\ast}\varepsilon_{\sR}^2(5+2\beta_{\mu})^2}\right] ,
   \nonumber\\ 
   a_{\ast}&=& (1-\varepsilon_{\sR}^2)(1-\varepsilon_{\sR}^2-\beta_{\mu}^2)
   , \nonumber\\
   b_{\ast}&=& (3+2\varepsilon_{\sR}^2)(1-\varepsilon_{\sR}^2+\beta_{\mu})-
   \varepsilon_{\sR}^2\beta_{\mu}(5+2\beta_{\mu}) . \nonumber
\end{eqnarray}  
In the limiting cases we have
\begin{eqnarray}
\chi_1 &\approx& \frac{12(1+\beta_{\mu})}{\beta_{\mu}(5+2\beta_{\mu})^2} , \ \  
\chi_2 \approx 0 \ \  \mbox{for } \ \varepsilon_{\sR} \rightarrow 0, \\
\chi_1 &\approx& 0 , \ \  \chi_2 \approx -\frac{8\beta_{\mu}}{(5+2\beta_{\mu})^2}
      \ \ \  \mbox{for } \ \varepsilon_{\sR} \rightarrow 1 .  \label{76}
\end{eqnarray}  
In Fig.~1b we plot for $|m|\ge 1$ and $\en_{\sR}\le 1$ the curves $\chi_1$
and $\chi_2$ versus $\en^2_{\sR}$ for a wide range of $\beta_{\mu}$. A
comparison of Figs.~1a and 1b shows that in the region {\bf III} with
$\en_{\sR}/m < -1/2$ LKHI will never appear. In the region {\bf I}
LKHI is possible for prograde waves, if $\beta_{\mu}<1/2$ and $\en_{\sR}^2 <
1-\beta_{\mu}(\beta_{\mu} + 3/2)$. With decreasing $\beta_{\mu}$ the solid
and dashed curves for the same $\beta_{\mu}$ are close to $\en_{\sR}\approx
1$. For retrograde waves LKHI is possible for strong gradients of
$\beta_{\mu}>1/2$ only in the range $-0.2 \le \en_{\sR}/m<0$. Here
$(\chi_2)_{\rm min}=-0.2$ is valid for $\en^2_{\sR}=1$ and $\beta_{\mu}=5/2$,
which follows from Eq.~(\ref{76}).

The total condition for the existence of spatially stable but temporarily 
unstable waves reads as follows: 
\be
   \chi_2 < \frac{\varepsilon_{\sR}}{m} < \min(\chi_1, \chi_3) .
\ee

\begin{figure}
\vspace{149mm}
\includegraphics{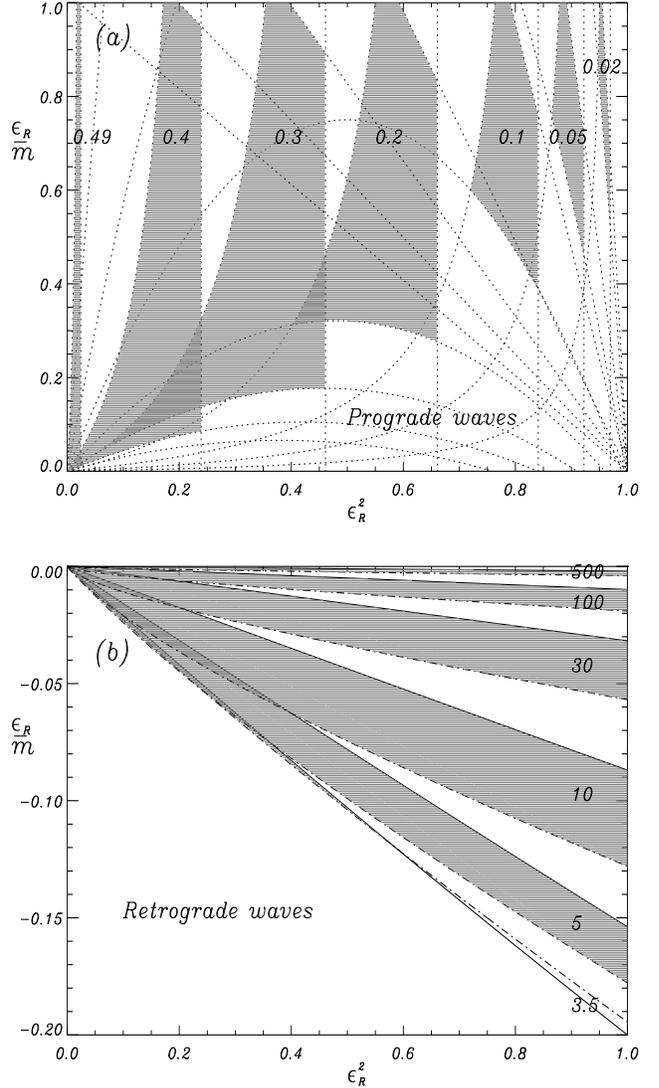}
\caption{ The wave instability areas (hatched) in the phase space from an
overlap of Figs.~1a and 1b. The labels in the areas are the $\beta_{\mu}$
values. Both prograde and retrograde waves $\beta_{\mu}$ are limited:
for prograde waves the instability is possible if $\beta_{\mu}<0.5$,
for retrograde waves the instability is possible if $\beta_{\mu}>3.5$.}
\end{figure}

Figs.~2a, b show the validity ranges of this condition for some typical
values of $\beta_{\mu}$. The hatched areas are places where LKHI is possible.
These figures are obtained by overlapping the Figs.~1a and 1b. For prograde
waves on both sides these hatched areas become very narrow: with decreasing 
$\beta_{\mu}$ the extent of the hatched area decreases and tends to the point
$\{\en_{\sR},m\}= \{1,1\}$. We will see that this is the solar case. The
hatched areas of LKHI disappear with decreasing $\beta_{\mu}$. This means we
have a lower limit $(\beta_{\mu})_{\rm min}$. 

For retrograde waves it is sufficient to write the condition as 
$\chi_2 < \en_{\sR}/m < \chi_3$. In Fig.~2a $\chi_2$ is the dashed
curve and $\chi_3$ is the solid curve. LKHI is possible if $\beta_{\mu}>3.5$
for $0\ge\en_{\sR}/m \ge -0.2$.

In Fig.~2 the hatched areas means that only these hatched areas for given
$\beta_{\mu}$ are possible, if LKHI takes place. Outside these hatched
areas regular solutions are impossible. The case without LKHI (neutral
oscillations) must be investigated separately. 

\subsection{Solar rotation profile}   

Let us consider at which places we might expect LKHI in the Sun.
Unfortunately, it is not clear how the core rotates. Nevertheless some
rotation gradients close to Sun's center might exist, and we could expect
LKHI there. It is known from helioseismology that the radiative interior
has a very small $\beta_{\mu}$, but the exact value is unknown. We have
better information on the rotation profile of the solar envelope, including
the tachocline. Helioseismology data may be described by different
approximate formulae. One of these is (Charbonneau et al. 1998)
\begin{eqnarray}
 \Omega (r, \theta) &=& \Omega_c+\frac{1}{2}\left[1+{\rm erf}(\Delta)\right]
 (\Omega_s(\theta)-
 \Omega_c) , \label{78} \\
\Omega_s &=& \Omega_{\rm eq}+c_1\cos^2\theta+c_2\cos^4\theta , \ \ 
\Delta\!=\!\frac{r\!-\!r_c}{w} ,  \nonumber\\
 \Omega_c/2\pi &=& 432.8\,{\rm nHz} ,  \Omega_{\rm eq}/2\pi = 460.7\,{\rm
 nHz} , \nonumber\\  
c_1 &=& -62.69\,{\rm nHz} , \ \  c_2 = -67.13\,{\rm nHz} ,  \nonumber
\end{eqnarray}  
where $r_c=0.713R_{\odot}$ is the radius at the bottom of the convective
zone, and $w=0.025R_{\odot}$ is the tachocline thickness. We can easily check
that our approximation $\tilde\Omega/\Omega_c \ll 1$ is always applicable, if
Eq.~(\ref{78}) is represented by $\Omega =\Omega_c + \tilde\Omega(r,\theta)$.
The maximum value $(\tilde\Omega/\Omega_c)_{\rm max}\approx 0.06$ for $\theta
=\pi/2$ (equator) is in the convection zone. From Eq.~(\ref{78}) we find
the latitudinal gradient of rotation
\be
   \beta_{\mu} = -\frac{1+{\rm erf}(\Delta)}{2\Omega}\sin^2\theta
   (c_1+2c_2\cos^2\theta) . \label{79}
\ee

\begin{figure}
\vspace{75mm}
\includegraphics{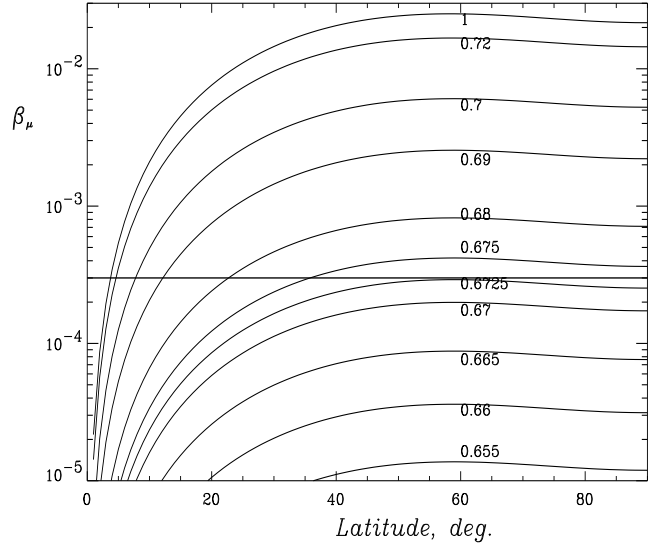}
\caption{ The local estimate of the logarithmic gradient of the solar
rotation frequency $\beta_{\mu}$ for real solar data from helioseismology
depending on the co-latitude and the radial distance (the labels are values
of $r/R_{\odot}$). The bold horizont is $\beta_{\mu}= 3\,10^{-4}$, 
above which the prograde waves become unstable (see next picture).}
\end{figure}

\begin{figure}
\vspace{75mm}
\includegraphics{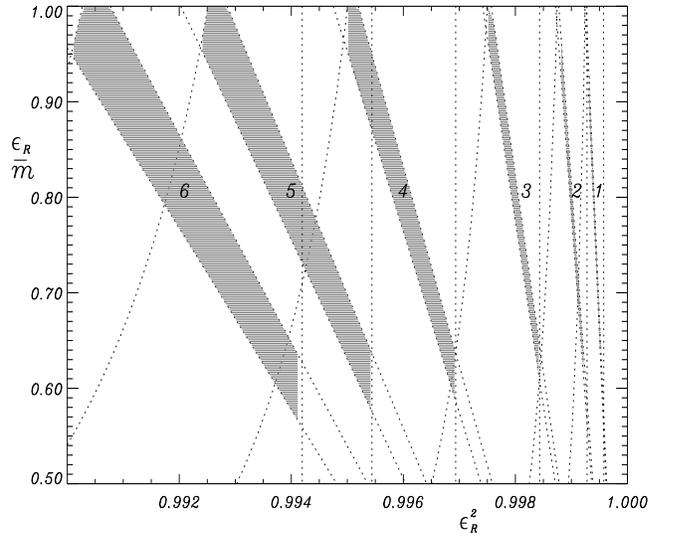}
\caption{ Enlarged part of Fig.~2a for the smallest gradients of rotation.
The labels 1, 2,..., 6 correspond to $\beta_{\mu}= 3\,10^{-4}, 
5\,10^{-4}, 1\,10^{-3}, 2\,10^{-3}, 3\,10^{-3}, 4\,10^{-3}$. Areas of
instability exist only if $\beta_{\mu} \ge 3\,10^{-4}$.} 
\end{figure}

Our supposition about $\beta_{\mu}\approx$ const and $\beta_r\approx$
const is based on the presentation $\tilde\Omega \approx \Omega_1(r)
+\Omega_2(\theta)$. We could receive such an approximate formula instead
of Eq.~(\ref{78}), but we need only local values of the gradients for
which Eq.~(\ref{79}) is acceptable. Using this formula we show in Fig.~3 the 
$\beta_{\mu}(\theta)$ dependence for different $r/R_{\odot}$. We
see that in the Sun $\beta_{\mu}\le 0.02$, and the maximum is in the 
photosphere. From Fig.~2 follows that the LKHI of retrograde waves is not
present in the solar case. LKHI of prograde waves in the Sun occurs in
the upper right corner of Fig.~2a which is enlarged
in Fig.~4. Here we see that the LKHI area disappears when $\beta_{\mu}<
3\,10^{-4}$. This boundary is located at the bold horizont in Fig.~3.
Thus the prograde waves become unstable in the Sun at those places where 
$3\,10^{-4} \le  \beta_{\mu} \le 2\,10^{-2}$. It means that LKHI is
possible in the area $r/R_{\odot}>0.6725$ which includes the greatest part of
the tachocline, the convective zone, and the photosphere. With increasing
$r$ the LKHI zone expands from middle to high latitudes. Figs.~2a and 4
show that LKHI is occurs at high frequencies ($\en_{\sR} \lesssim 1$)
and in global scales ($\en_{\sR}/m > 0.5$). Considering that $m\ne 0$ is an
integer we get $m=1$.

However, our quantitative analysis of LKHI is based on the general Riemann
scheme of Hein's equation, which is valid only if the middle singular point
$x=a$ is far from the other edges at $x=0$ and $x=1$. Thus, the limiting
cases $\en_{\sR} \rightarrow 0$ and $\en_{\sR} \rightarrow 1$ (the latter is
more important for solar LKHI) should be considered separately. In these
limiting cases the regularity condition Eq.~(48) may be changed, and the
curve in Fig.~4 limiting the instability areas from below may be shifted. In
this case LKHI with higher $m$-modes should be expected. 

\section{The low-frequency waves}   

After the qualitative analysis of Heun's Eq.~(\ref{69}) we can start a
quantitive analysis. Note that the qualitative conclusions drawn above are
valid for the more general Eq.~(\ref{59}) with $\Lambda$ term. Heun's
equation with four singularities in the general case is solved by a series
of hypergeometric Gauss functions. A similar task has been considered for the
damping of MHD waves at resonance levels by Dzhalilov \& Zhugzhda (1990). We
will start to study Eq.~(\ref{69}) for some simple limiting cases. At
high frequencies ($\en^2_{\sR}\approx 1$, when LKHI is acting in the Sun) and
at low frequencies  ($\en^2_{\sR} \ll 1$, when the waves are stable against
LKHI in the Sun) Heun's equation is strongly simplified. In these cases the
singular level $x=a$ is shifted either to the pole or to the equator. For
both cases solutions are expressed by one hypergeometric function.

In the present work we consider particularly the second case. Let
$\en^2_{\sR} \ll 1+\beta_{\mu}$. Then we have $a\approx 1$ and $h\approx 1$.
Eq.~(\ref{69}) is now the hypergeometric equation:
\be
x(1-x)Y^{\prime\prime}+[\gamma-(\tilde{\alpha}+\tilde{\beta}+1)x]
Y^{\prime}-\tilde{\alpha}\tilde{\beta}Y=0 ,
\ee
where all parameters are defined by Eqs.~(\ref{67}, \ref{68'}, \ref{71}).
In these definitions $\en^2_{\sR}$ should set to zero. Then the $\Theta$-part
of the pressure perturbations, Eq.~(\ref{66}), is expressed by two
Gaussian hypergeometric functions:
\be
\Theta = C_1x^{\sigma_1}Y_1(x) + C_2x^{\sigma_2}Y_2(x) . \label{81}
\ee
\be
{\rm Here \ }Y_{1, 2}(x)=F(\tilde{\alpha}, \tilde{\beta}; \gamma;
x)|_{\sigma=\sigma_{1, 2}} , 
\ee
and $C_{1, 2}$ are arbitrary constants. This general solution includes LKHI 
for larger $\beta_{\mu}$ too. This could be realized perhaps in other,
younger stars. For the Sun we have $\beta_{\mu} \ll 1$. We will finish this
paper by considering in detail the more popular case when rotation is
uniform, $\beta_{\mu} =0$.

\begin{figure}
\vspace{75mm}
\includegraphics{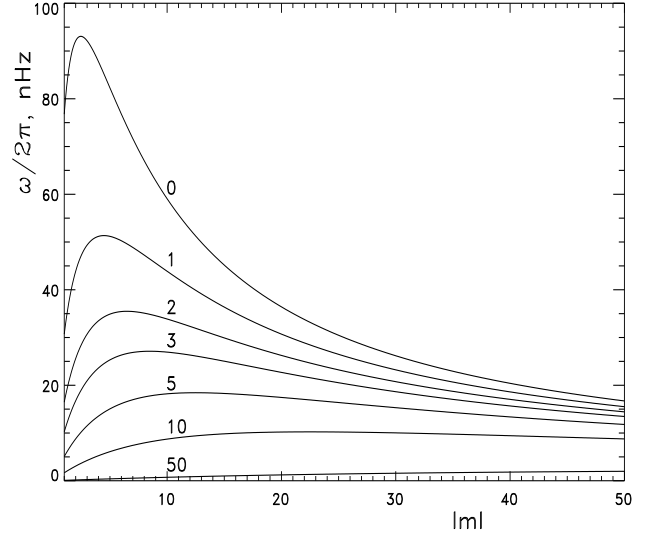}
\caption{The spectrum $\omega(l,m)$ of low-frequency retrograde modes.
The numbers at the curves are the $l$ values. For the calculation of the
spectrum, Eq.~(\ref{89}), $\Omega=\Omega_{\rm eq}=460.7$ nHz is used.}
\end{figure}

\subsection{Rigid rotation case}

Using the conditions $\beta_{\mu}=0$ and $\en^2_{\sR} \ll 1$ the
parameters in the solution Eq.~(\ref{81}) are strongly simplified. Because
$2\sigma=\pm |m|$ only a regular solution at the pole ($x=0$) will be left.
In the standard definitions of hypergeometric functions (Abramowitz \& Stegun
1984) we have 
\begin{eqnarray}
\Theta &=& Cx^{|m|/2}F(a, b; c; x) ,\label{83} \\
a &=& \frac{1}{2}\left(-\frac{1}{2}+|m|+q\right) , \ 
b=\frac{1}{2}\left(-\frac{1}{2}+|m|-q\right) ,   \nonumber \\
c &=& 1+|m| , \,\,\, \ q=\sqrt{\frac{1}{4}-\frac{m}{\varepsilon_{\sR}}} ,
\ \ {\rm and} \ \  C ={\rm const} .\nonumber 
\end{eqnarray}
Note that $c-a-b=3/2$. The analytical continuation of
the solution Eq.~(\ref{83}) to the equator, $x\to 1$, gives
\begin{eqnarray}
\Theta &=& Cx^{|m|/2}\left[AF(a, b; -\frac{1}{2}; 1-x)+\right. \nonumber \\
&+&\left.B(1-x)^{3/2} F(c-a, c-b; \frac{5}{2}; 1-x)\right] . \label{84}
\end{eqnarray}
Here the continuation coefficients are (Abramowitz \& Stegun 1984)
\begin{eqnarray}
A &=& \frac{\Gamma(c)\Gamma(3/2)}{\Gamma(c-a)\Gamma(c-b)}=F(a, b; c; 1) ,
\label{85'} \\
B&=&\frac{\Gamma(c)\Gamma(-3/2)}{\Gamma(a)\Gamma(b)} . \label{85}
\end{eqnarray}
Eqs.~(\ref{83}, \ref{84}) mean that pressure is limited everywhere in the
hemisphere. Now let us consider the velocity components. Putting
Eq.~(\ref{83}) into Eq.~(\ref{63}) we get
\begin{eqnarray}
V_{\theta}&=&\pm\frac{CQ\varepsilon_{\sR}^2}{r\rho_0 i\omega}
\frac{x^{|m|/2}}{\sqrt{x(1-x)}}
\left[\left(\frac{m}{\varepsilon_{\sR}}-|m|\right)F(a, b; c;
x)-\right. \nonumber \\
&-&\left. x\frac{2ab}{c} F(a+1, b+1; c+1; x)\right] .\label{86}
\end{eqnarray}
Taking into account that $F(a,b;c;0)=1$, we receive from the regularity of
$V_{\theta}$ at $x=0$ that $|m|\ge 1$. Axially-symmetric waves
$m=0$ cannot be formed. The continuation of the solution Eq.~(\ref{86}) to
the equator, $x\to 1$, gives  
\begin{eqnarray}
V_{\theta} &=& \pm\frac{CQ\varepsilon_{\sR}^2}{r\rho_0 i\omega}x^{(|m|-1)/2}
\left(\frac{A}{\sqrt{1-x}}{\cal L}_1+B{\cal L}_2\right) , \label{87} \\
{\cal L}_1 &=& \left(\frac{m}{\varepsilon_{\sR}}-|m|\right)F(a, b; -\frac{1}{2};
1-x)-\nonumber\\
&&-x4abF(a+1, b+1; \frac{1}{2}; 1-x) , \nonumber\\
{\cal L}_2 &=&
\left[\left(\frac{m}{\varepsilon_{\sR}}\!-\!|m|\right)(1\!-\!x)\!+\!3x\right]
F(c\!-\!a, c\!-\!b; \frac{5}{2}; 1\!-\!x)+ \nonumber\\
&+&\frac{4}{5}x(1\!-\!x)(c\!-\!a)(c\!-\!b)F(c\!-\!a+1, c\!-\!b\!+\!1;
\frac{7}{2}; 1\!-\!x) . \nonumber
\end{eqnarray}
As for $x=1$ the functions ${\cal L}_1$ and ${\cal L}_2$ are limited, we have
the relation $A\equiv 0$ from the regularity requirement. Using a property
of the gamma-functions (its presentation as an infinite product) in
Eq.~(\ref{85'}) we get the condition of quantization
\begin{equation}
q^2 = \left(2l+|m|+\frac{5}{2}\right)^2, \, \, \, l=0, 1, 2, ... \label{88}
\end{equation}
From here we get the simple dispersion relation
\be
\frac{\varepsilon_{\sR}}{m}=\frac{1}{m}\frac{\omega}{2\Omega}=-\frac{1}
{(2+2l+|m|)(3+2l+|m|)} . \label{89}
\ee
Since $\en_{\sR}/m < 0$ for any $l\ge 0$, only retrograde modes (as seen in
the rotating frame) are possible. 

\begin{figure}
\vspace{75mm}
\includegraphics{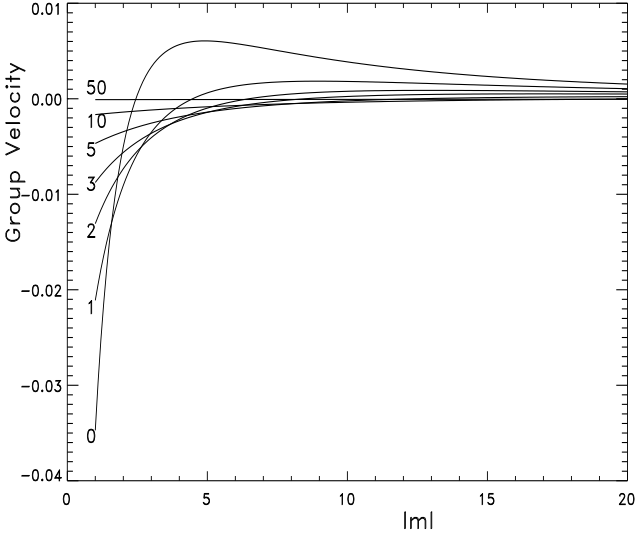}
\caption{The normalized group velocity $v_g/2\Omega r_0\sin\theta_0$ as a 
function of the azimuthal numbers for given degrees $l$.}
\end{figure}

\subsubsection{Spectrum of retrograde modes}

The new dispersion relation Eq.~(\ref{89}) completely differs from the
dispersion relation of the almost toroidal r-modes. Their dispersion relation
can be derived from Eq.~(\ref{89}) if we formally set $l_s = 2(1+l)+|m|$.
Then
\begin{equation}
\omega=-2\Omega\frac{m}{l_s(1+l_s)} . \label{90}
\end{equation}
However, here the degrees $l_s$ are functions of $m$. Due to the coupling of
the modes in our case the eigenfuncions can never be expressed by the
associated Legendre functions $P^m_{l_s}$. For the $r$-modes the
axi-symmetric modes ($m=0$) with $\omega=0$ are possible, while our case
rejects this case. Unlike the $r$-modes the spectrum Eq.~(\ref{89}) has a
maximum at $m^2 = m^2_0 =2(1+l)(3+2l)$.  If $l \gg 1$ we have
$|\en_{\sR}|_{\rm max} \approx 1/8l$. The spectrum of retrograde waves is
shown in Fig.~5. Rossby waves in  geophysics (Pedlosky 1982) have a similar
spectrum. Using Eq.~(\ref{89}) we may define the local phase and group
velocities (at fixed latitude $\theta_0$ and radial distance $r_0$) in the
azimuthal plane
\begin{eqnarray}
v_{\rm ph} &=& -\frac{2\Omega r_0\sin\theta_0}{(2+2l+|m|)(3+2l+|m|)} ,
\label{91} \\
v_{\rm gr} &=&
2\Omega r_0\sin\theta_0\frac{m^2-2(1+l)(3+2l)}{(2+2l+|m|)^2(3+2l+|m|)^2} .
\label{92}
\end{eqnarray}
We see that the group velocity changes its sign at the maximum of the
spectrum, where $m^2=m^2_0$. Fig.~6 shows that the $l=0$ mode has a maximum
group velocity. Long waves carry energy opposite to the rotation
direction, while a packet of short waves is carried in rotation direction.
The facts that the frequencies $\omega(m)$ have a maximum (two different
$|m|$ correspond to the same $\omega$) and at the maximum of
$\omega=\omega_{\rm max}$ the group velocity changes the direction hints at
the existense of two branches of oscillations. Solving Eq.~(\ref{89}) for
$|m|$ we get these branches. Let $\omega>0$ and $m=-|m|$. Then
\begin{eqnarray}
|m|&=&\frac{1}{2\varepsilon_{\sR}}(w_1\mp\sqrt{w_2})=m_{1, 2} \ ,\label{93}\\
w_1 &=& 1-\varepsilon_{\sR}-4\varepsilon_{\sR}(1+l) , \nonumber\\
w_2 &=& (1-\varepsilon_{\sR})^2-8\varepsilon_{\sR}(1+l) .\nonumber
\end{eqnarray}

\begin{figure}
\vspace{75mm}
\includegraphics{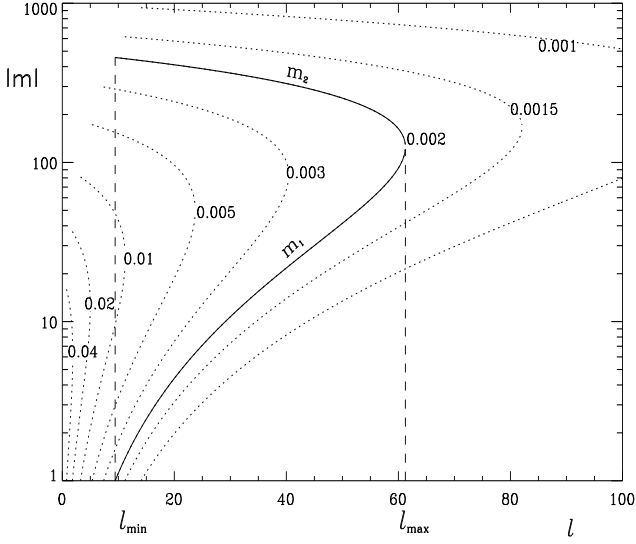}
\caption{The possible domain of the existence of eigenmodes for given
Rossby numbers $\en_{\sR}$ (the numbers at the curves). For example,
the case $\en_{\sR}=0.002$ (close to the 22-year modes) is emphasized.
All possible values of the azimuthal numbers ($m=m_1$ and $m=m_2$) are on
this  curve for given discrete $l$ in the range $l_{\rm min}\le l \le
l_{\rm max}$ (see text). If  $\en_{\sR}>1/12$, the eigenmodes disappear.}
\end{figure}

From here we get an upper limit for $l$ if $\en_{\sR}$ is given: $l\le l_{\rm
max}= (1-\en_{\sR})^2/8\en_{\sR} -1$. For such degrees of $l$ the azimuthal
numbers are also limited: $m_1\le|m|\le m_2$. For  $\en_{\sR}\to 0$ we have
$l_{\rm max}\to 1/8\en_{\sR}$, $m_2\to 1/\en_{\sR}$, and $m_1\to 0$. However,
considering the regularity of the solutions Eq.~(\ref{86}), we must take
$m_1\ge 1$. From here we get the lower limit of $l$:
\be
l \ge l_{\rm min} =
\frac{1}{4}\left(\sqrt{1+\frac{4}{\varepsilon_{\sR}}}-7\right) .\label{94}
\ee
Since $l\ge 0$, we get $\en_{\sR}\le 1/12$ from Eq.~(\ref{94}).  For
$\en_{\sR}= 1/12$ we have $l_{\rm min}=0$. Thus the eigenmodes exist for
$\en_{\sR}\le 1/12$ if $l_{\rm min}\le l\le l_{\rm max}$ and $m_1\le |m|\le
m_2$. For $l=l_{\rm max}$ we have $m_1=m_2$, and $m_1=1$ for $l=l_{\rm min}$.
This situation is shown in Fig.~7 for different values of
$\en_{\sR}$. A decrease of the frequency decreases the domain of the
existence of the modes.

Setting Eq.~(\ref{93}) into  Eqs.~(\ref{91}, \ref{92}) gives
\begin{eqnarray}
\frac{v_{\rm ph}^{\pm}}{2\Omega r_0\sin\theta_0} &=&
-\frac{w_1\pm \sqrt{w_2}} {4(1+l)(3+2 l)} , \label{95}\\
\frac{v_{\rm gr}^{\pm}}{2\Omega r_0\sin\theta_0} &=&
-\frac{w_2 \pm w_1 \sqrt{w_2}} {4(1+l)(3+2 l)} . \label{96}
\end{eqnarray}

\begin{figure}
\vspace{75mm}
\includegraphics{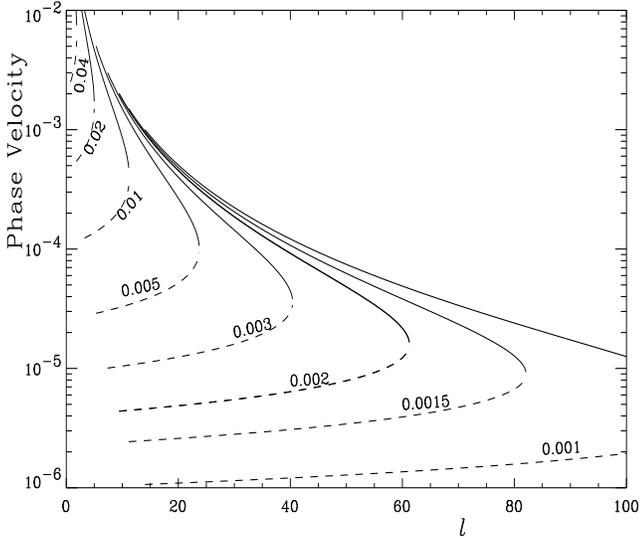}
\caption{The phase velocities of fast (solid) and slow (dashed curves)
modes versus $l$. The numbers at the curves are $\en_{\sR}$ values,
similar to Fig.~7. Here the velocities are normalized to ($-2\Omega r_0\sin
\theta_0$). Every curve is restricted to the range $l_{\rm min}\le l\le
l_{\rm max}$.}
\end{figure}

Here $v_{\rm ph}^+$ is the phase velocity of the fast modes and $v_{\rm
ph}^-$ that of the slow modes. For $l=l_{\rm max}$ we have $v_{\rm
gr}^{\pm}=0$ and $v_{\rm ph}^+=v_{\rm ph}^-$. In Eq.~(\ref{93}) and in Fig.~7
the $m_1$ branch corresponds to the fast mode, but $m_2$ to the slow modes,
since $m_2\ge m_1$. In Fig.~8 the normalized phase velocities (with inverse
sign) for the selected values of $\en_{\sR}$ in Fig. 7 are shown versus $l$.
Both branches are retrograde modes ($v_{\rm ph}^{\pm}<0$). Using $\Omega
R_{\odot} \sim 2$ km/s for the Sun, we get from Fig.~8 very slow phase
velocities. The fast wave velocity (solid lines) depends more strongly on
$l$. With increasing $\en_{\sR}$ both branches are accelerated.

In Fig. 9 the group velocities are presented in the same way. For fast waves
the group velocity is always parallel to the phase velocity ($v_{\rm
gr}^+<0$), while for the slow waves we have the opposite behavior $v_{\rm
gr}^->0$. Slow modes packets carry off energy in the rotation direction.
Always $|v_{\rm gr}^+|>|v_{\rm gr}^-|$ is valid. With deceasing
$\en_{\sR}$ the range [$l_{\rm min}, l_{\rm max}$] is shifted to the
right-hand side, and it is seen in Fig.~9 that $v_{\rm gr}^-$ for such low
$\en_{\sR}$ is almost zero. 

Note that $m=l$ modes are always fast modes.

\subsubsection{The eigenfunctions}   

Taking into account the quantization condition Eq.~(\ref{88}) in the 
solutions Eqs.~(\ref{84}, \ref{87}, \ref{62}, and \ref{64}) we obtain
the eigenfunctions. Turning from  complex velocities into the real 
displacements, $\vec{V}=-i\omega\vec\xi$ (recall that $\vec{V}$ is the
velocity seen in the rotating frame), we get  
\begin{eqnarray}
p' &=&
Q(r)\Theta(\theta)\cos\left[m\left(\phi-2\frac{\varepsilon_{\sR}}{m}\Omega
t\right)\right] ,\label{97} \\
\xi_{\theta} &=& \xi_{\theta}^{\star}\cos\left[m\left(\phi
-2\frac{\varepsilon_{\sR}}{m}\Omega t\right)\right] ,\label{98} \\
\xi_{\phi} &=& \xi_{\phi}^{\star}\sin\left[m\left(\phi
-2\frac{\varepsilon_{\sR}}{m}\Omega t\right)\right] ,\label{99} \\
\xi_{r} &=& \xi_{r}^{\star}\cos\left[m\left(\phi
-2\frac{\varepsilon_{\sR}}{m}\Omega t\right)\right] .\label{100} 
\end{eqnarray}
Here the amplitude functions are
\begin{eqnarray}
 \xi_{r}^{\star} &=&
 \frac{1}{\omega^2-N^2}\frac{\Theta}{\rho_0}\frac{dQ(r)}{dr} , \label{102} \\
 \xi_{\theta}^{\star} &=& \frac{Q(r)}{4\Omega^2r\rho_0}\xi_{\theta A}(\theta) ,
 \label{103} \\
 \xi_{\phi}^{\star}  &=& \frac{Q(r)}{4\Omega^2r\rho_0}\xi_{\phi A}(\theta) .
 \label{104}
\end{eqnarray}

\begin{figure}
\vspace{75mm}
\includegraphics{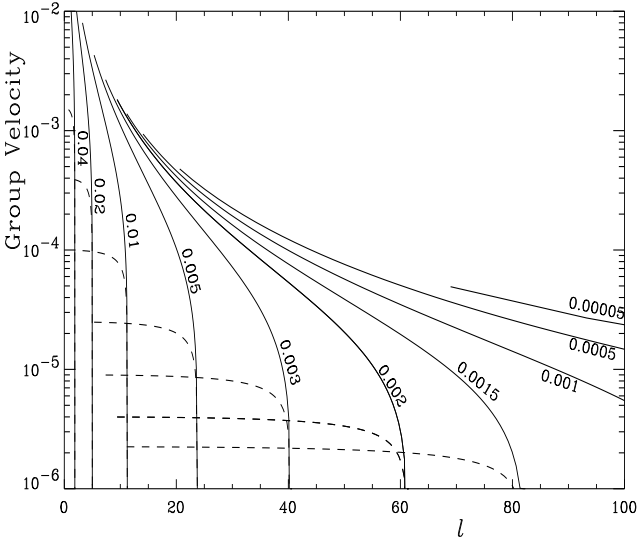}
\caption{Absolute values of the group velocities of fast (solid) and slow 
(dashed curves) modes normalized to $2\Omega r_0\sin\theta_0$ versus $l$ for
selected $\en_{\sR}$. For fast waves $v_{\rm gr}^+<0$, for slow modes
$v_{\rm gr}^->0$. For $l=l_{\rm max}$ we have $v_{\rm gr}^+=v_{\rm gr}^-=0$.}
\end{figure}

The amplitudes are expressed by Jacobi's polynomials:
\begin{eqnarray}
\Theta &=& B(\sin\theta)^{|m|}\cos^3\theta \,F_1(\theta) , \label{105}\\
\xi_{\Theta A} &=&  B(\sin\theta)^{|m|-1}\cos^2\theta
\left(\frac{m}{\varepsilon_{\sR}}F_1-\frac{4lk}{5}\sin^2\theta F_2\right) ,
\label{106} \\
\xi_{\phi A} &=& -\frac{B}{\varepsilon_{\sR}}\frac{2lk}{5}
\sin(2\theta)\cos^2\theta(\sin\theta)^{|m|}F_2(\theta) , \label{107}
\end{eqnarray}
\begin{equation}
F_1 = F\left(-l, k; \frac{5}{2}; \cos^2\theta\right)=  \label{108}
\end{equation}
 \[ \ \ \ \ = (-1)^l \frac{l!\Gamma(5/2)}{\Gamma(l+5/2)} P_l^{(|m|,
 3/2)}(\cos 2\theta)=  \]
 \[ \ \ \ \ = \sum_{j=0}^{l} \frac{(-l)_j(k)_j}{(5/2)_j\,j!}(\cos\theta)^{2j}
 , \] 
\begin{equation}
F_2 = F\left(1-l, 1+k; \frac{7}{2}; \cos^2\theta\right)=  \label{109}
\end{equation}
\[ \ \ \ \ = (-1)^{l-1}
\frac{(l-1)!\Gamma(7/2)}{\Gamma(l+5/2)}P_{l-1}^{(1+|m|, 5/2)}(\cos 2\theta) =
\nonumber \] \[ \ \ \ \ = \sum_{j=0}^{l-1} \frac{(1-l)_j(1+k)_j}{(7/2)_j\,
 j!}(\cos\theta)^{2j} .\] 
\begin{equation}
   k=\frac{5}{2}+|m|+l , \ \ (f)_j=\frac{\Gamma(f+j)}{\Gamma(f)} ,
\end{equation}
and $F_2\equiv 0$ if $l=0$.
The eigenfunctions of the retrograde low-frequency modes
Eqs.~(\ref{97}--\ref{100}) are written for the hemisphere
$0\le\theta\le\frac{\pi}{2}$. In the other hemisphere we have to change
$\cos\theta$ to $|\cos\theta|$ and to change the sign of $\xi_{\theta}$. We
put $C=1$ because all eigenfunctions are  multiplied by the solution of the
$r$-Eq.~(\ref{57}). Changing the sign of $m$ the function $\xi_{\phi}$ keeps
its sign because $\en_{\sR}/m<0$. It follows from Eqs.~(\ref{108},
\ref{109}) that the eigenfunctions are represented by a limited number of
trigonometric functions. The number of terms is defined by $l$ in the range
$l_{\rm min}\le l \le l_{\rm max}$. This is qualitatively different from the
presentation of the rotation-mode eigenfunctions by infinite series of
spherical harmonics.

\begin{figure*}
\vspace{149mm}
\includegraphics{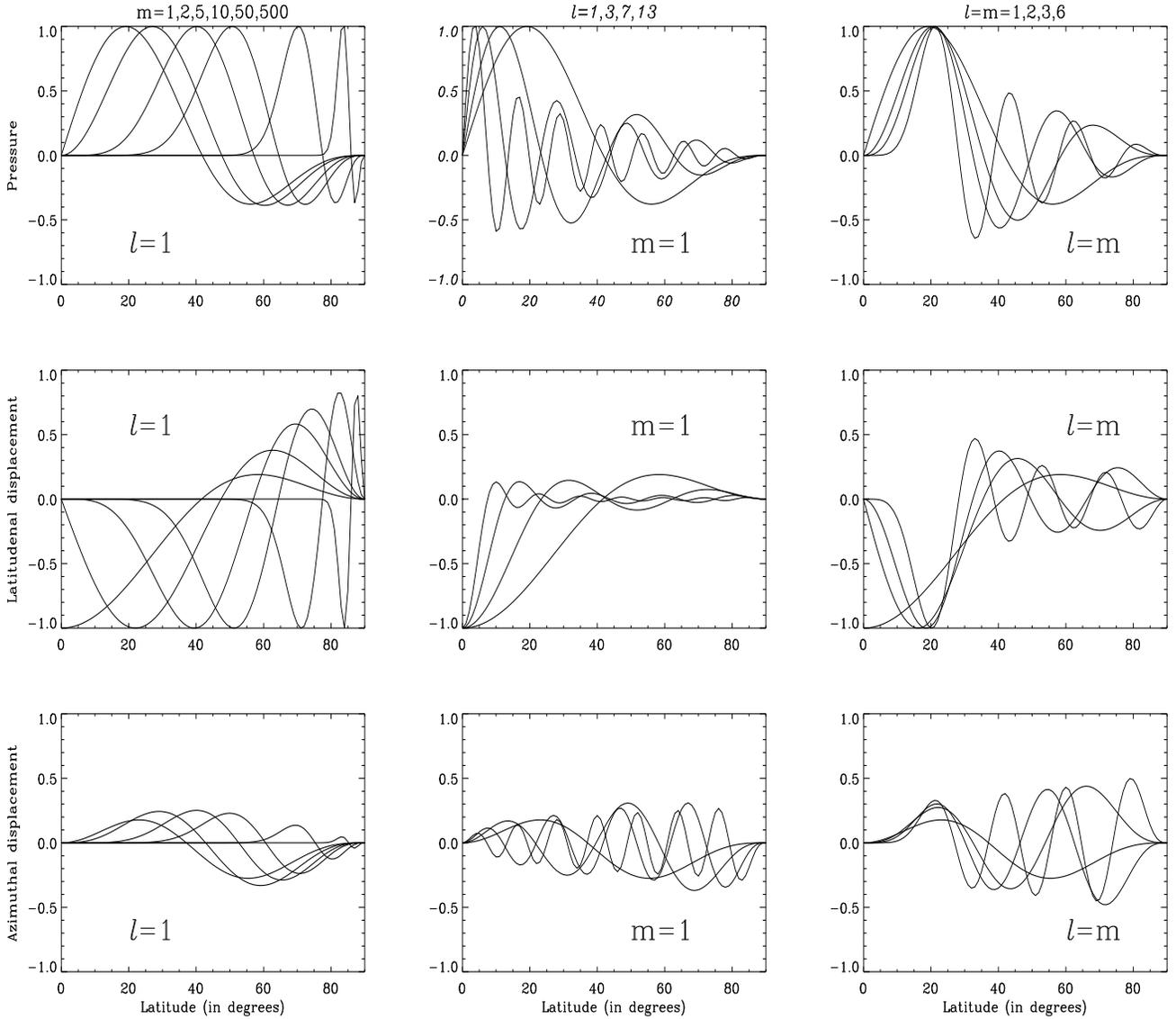}
\caption{The amplitudes of eigenfunctions versus latitude for given pairs
($l,|m|$), Eqs.~(\ref{105})--(\ref{107}). The pressure row shows the
$\Theta(\theta)$ function normalized to its own maximum. From left to right
all curves are related to the wave numbers given above. The middle and last
rows are similar to the first row, but show now the latitudinal ($\xi_{\theta
A}$) and azimuthal ($\xi_{\phi A}$) amplitude functions, normalized to the
maximum of ($\xi_{\theta A}$).}
\end{figure*}

Let us consider some particular cases.
\begin{itemize}
 \item[a)] Pole ($\theta=0$): taking into account $B F_1=1$ we have
 $\Theta= p'=\xi_r=\xi_{\phi}=0$. $\xi_{\theta}=0$ if $|m|>1$. For  $|m|=1$ 
we have $\xi_{\theta A}=-2(3+2l)(2+l)$.

\item[b)] Equator ($\theta=\frac{\pi}{2}$): here $F_1=F_2=1$ and
$p'=\xi_{\theta}= \xi_{\phi}=0$.

\item[c)] $l=0$ case: then $B=F_1=1$, $F_2=0$, 
$$
\Theta=(\sin\theta)^{|m|}\cos^3\theta ,
$$
$$\xi_{\theta A}=-(2+|m|)(3+|m|)\cos^2\theta(\sin\theta)^{|m|-1} ,
$$ 
and $\xi_{\phi}=0$ (in our $\en^2_{\sR} \ll 1$ limit).

\item[d)] $l \gg |m|$, $|m|$ is not large, and $\theta\ne 0,\frac{\pi}{2}$
as well. 
\end{itemize}
The latter case is important because the range $l_{\rm min}\le l \le
l_{\rm max}$ is large for small $\en^2_{\sR}$. Remember that for
$\en_{\sR}=1/12$ we have $l_{\rm min}=0$. Using the asymptotic formula for
Jacobi's polynomial (or the hypergeometric function) of large degree we
have
\begin{eqnarray}
\Theta &\simeq& \frac{|m|!}{\sqrt{\pi l}}
\frac{\cos\lambda \cos\theta}{l^{|m|}\sqrt{\sin\theta}} , \\
\xi_{\theta A} &\simeq & \frac{|m|!}{\sqrt{\pi l}}
\frac{\cos\lambda}{l^{|m|}(\sin\theta)^{3/2}}\!
\left(\frac{m}{\varepsilon_{\sR}}\!+\!\frac{14}{5}k\tan\theta
\tan\lambda\right) ,
\\
\xi_{\phi A} &\simeq & \frac{14}{5}\frac{|m|!k}{\varepsilon_{\sR}\sqrt{\pi
l}} \frac{\sin\lambda}{l^{|m|}\sqrt{\sin\theta}} .
\end{eqnarray}
\[ \lambda=\left(2l+\frac{5}{2}+|m|\right)\theta
-\frac{\pi}{2}\left(|m|+\frac{1}{2}\right) . \]
These extremely simple asymptotic formulae for the eigenfunctions might be
used in most cases. The formula for the case of large $l$ with large $|m|$
also could be found, e.g. in the book of Bateman \& Erd\'{e}lyi (1953).

In Fig.~10 for some typical selected ($l,m$) pairs the amplitude functions,
Eqs.~(\ref{105}--\ref{107}), are shown as function of $\theta$. The first row
is the pressure $\Theta(\theta)$ function normalized to its maximum. The
first $l=1$ window represents $\Theta$ for different $|m|$ (increasing $|m|$
from left to right). As the eigenfunctions are multiplied by a
$\sin^{|m|}\theta$ factor, the amplitudes are strongly suppressed around the
pole. Increasing $|m|$ for a given $l$ shifts the  maximums toward the
equator. $l$ is the surface node number of the $\Theta(\theta)$ function.
Increasing $l$ for given $|m|$ (the second window of pressure with
$m=1$ in Fig.~10) contrarily suppresses the amplitudes around the equator,
and the maximum is shifted toward the pole. $l=|m|$  is the equilibrium
case. In the third window of pressure the balance latitude  with a maximum
amplitude is defined by $\theta=20^{\circ}$ for all $l=|m|$ modes.

The second and third rows of Fig.~10 are the latitudinal 
($\xi_{\theta A}$) and azimuthal ($\xi_{\phi A}$) eigenfunction amplitudes, 
respectively, normalized to the maximum of $\xi_{\theta A}$, see 
Eqs.~(\ref{103}, \ref{104}). The latitudinal amplitude behavior is
similar that of the pressure. The azimuthal amplitudes are smaller than the
latitudinal amplitudes, but with a change of $l$ a redistribution of the
amplitudes will not take place. $\xi_{\phi A}$ has practically the same
amplitude at all latitudes and for all $l$.

From Fig.~10 follows that we can expect an interesting behavior of the
eigenfunction amplitudes, when both $l$ and $|m|$ are large. A suppression
from two sides may evoke a concentration of wave energy in narrow latitudinal
bands. For example, this is the case for the 22-year solar mode. 

\begin{figure}
\vspace{75mm}
\includegraphics{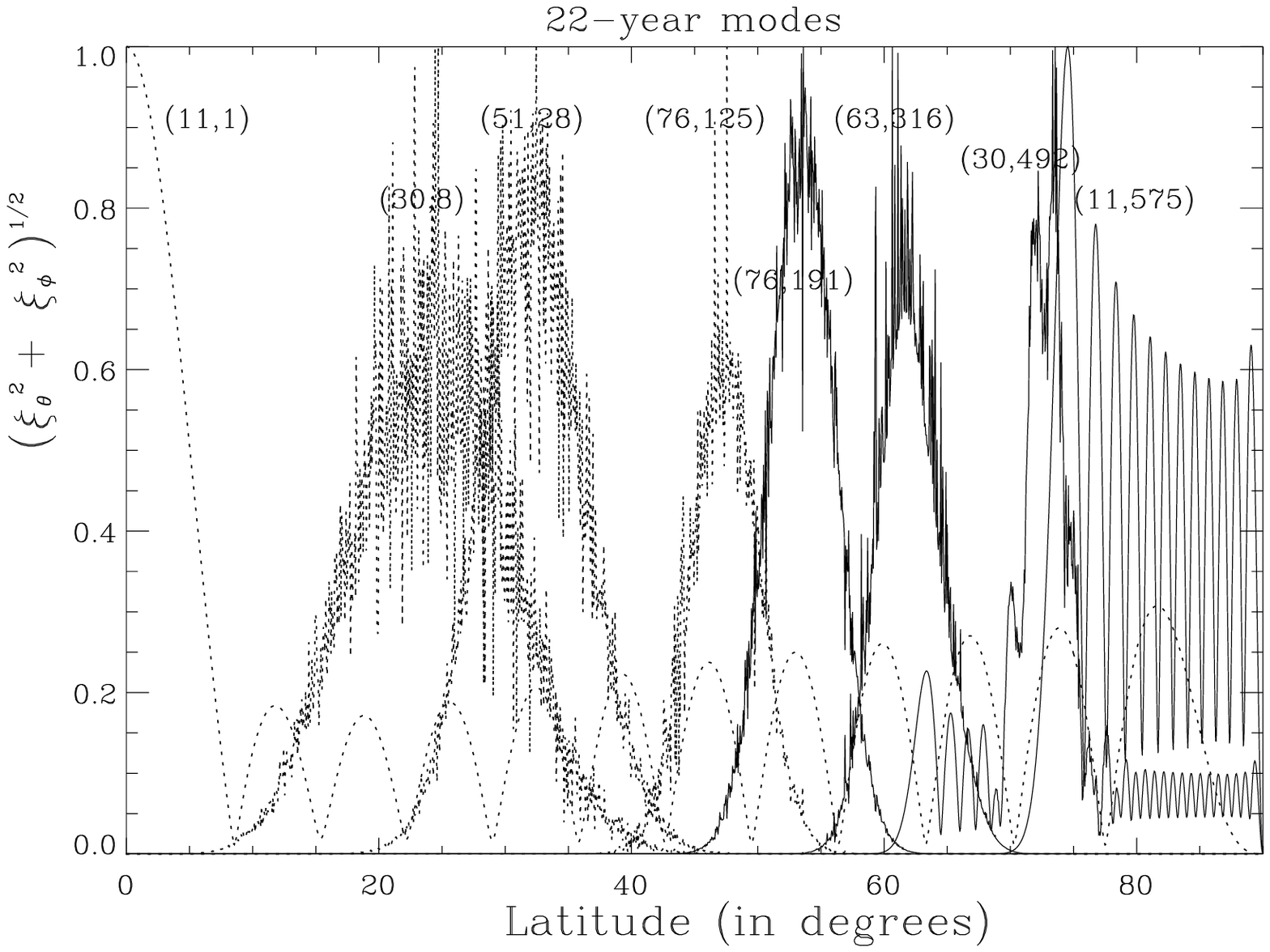}
\caption{The normalized energy density of the 22-year fast (dashed) and
slow (solid) modes versus co-latitude. The numbers at the curves are
($l,|m|$)  pairs taken from Table 1.}
\end{figure}

For the 22-year modes we take $\omega_{22}=2\pi$(1.441 nHz), for
which $\en_{\sR}\approx 0.0016$. Then we derive from Eq.~(\ref{94}) the
limiting values of the integer $l$, $11\le l\le 76$. For all $l$ in
this range we find from Eq.~(\ref{93}), rounding off, integer azimuthal
numbers $m_1$ and $m_2$ for the fast and slow modes, respectively. Putting
these integer numbers into Eq.~(\ref{89}) we find the deviation
$\delta=\omega_{22}-\omega$ from the central frequency due to the integer
azimuthal numbers. The results are given in Table 1 in Appendix A. It is
seen that the fast modes with low $l$ have larger deviations. This table
includes all possible ($l,|m|$) pairs which  correspond to the 22-year
period. For some example pairs of ($l,|m|$) we plot in Fig.~11 the
latitude dependence of the the quantity $(\xi^2_{\theta}+
\xi^2_{\phi})^{1/2}$, averaged over the wave period, which characterizes the
energy density of the modes. The hemisphere is divided into two equal parts:
slow modes are located around the equator (solid lines), fast modes are
concentrated around the pole. Each ($l,|m|$) pair is located in a narrow 
latitudinal band. Note that the slow modes (the group velocity of which is in
the rotation direction) with sunspot-like spatial scales are at latitudes of
$30-40^o$ from the equator. 

\subsubsection{Flow patterns}   

The eigenfunctions Eqs.~(\ref{97}--\ref{100}) allow us to discuss the
flow character produced by the waves, even if the solution of the radial
equation $Q(r)$ is unknown. Excluding from these equations the time-dependent
phase we can receive the trajectory equations of the fluid elements. In the
meridional $(\vec r,\vec\theta)$ plane we have   
\be
 \frac{\xi_{\theta}}{\xi_{r}}=\varepsilon_{\sR}^2\frac{Q(r)\psi(r)}
 {r\,Q'(r)} \frac{\xi_{\theta A}}{\Theta}=\tan(\alpha_r) . \label{114}
\ee
It follows from here that the motion in the meridional plane is linear.
The poloidal displacement vector in the meridional plane is inclined to 
the radius by an angle $\alpha_r$. For $\alpha_r\approx\frac{\pi}{2}$
we have meridional flows and for $\alpha_r\approx 0$  radial flows.
However, since $\xi_{\theta}/\xi_r\sim 1/\sin(2\theta)$, for any $r$ the
motions around the pole and around the equator will be almost meridional 
($|\xi_{\theta}| \gg |\xi_r|$). We see from Eq.~(\ref{114}) also that
the direction of the flow vector will be changed to $\pi/2$ if we pass along
the radius from the node of the $Q(r)$-function to $Q'(r)$. Note that every
node of $Q'(r)$ is located between the neighbouring two nodes of $Q(r)$. For
large radial numbers (such orders are expected) these nodes are located
close to each other. Then we have a complicated motion in the meridional
plane.

At the surface of the cone over $(\vec r,\vec\phi)$ the trajectory of each
fluid element is an ellipse around the equilibrium point. The cone
displacement equation is
\begin{equation}
\frac{\xi_{\phi}^2}{\xi_{\phi}^{\star 2}}+\frac{\xi_{r}^2}{\xi_{r}^{\star
2}}=1. \label{115}
\end{equation}
The ratio of the semiaxes $\xi_{\phi}^*$ and $\xi_r^*$ 
\be
\frac{\xi_{\phi}^{\star}}{\xi_{r}^{\star}}=-\varepsilon_{\sR}
\frac{Q(r)\psi(r)}{rQ'(r)}\frac{4lk}{5}\frac{F_2}{F_1}\sin\theta 
\ee
means that close to the axis of rotation the flow is directed along this
axis. On the bigger cone through the equator, where $F_1=F_2=1$, the motion
is mainly azimuthal since $lk>1$. An exception are in this case the nodes
of the $Q(r)$-function at the radius. For large $l \gg 1$ the ratio
$\xi_{\phi}^*/\xi_r^*\sim \tan(2l\theta)/\cos\theta$ means that the character
of the flows is changing at middle latitudes.

At the surface of any sphere with a radius $r$ the motion of the fluid
elements is on trajectories of the ellipse
\be
\frac{\xi_{\theta}^2}{\xi_{\theta}^{\star 2}}
+\frac{\xi_{\phi}^2}{\xi_{\phi}^{\star 2}}=1.
\ee
Here the ratio of the semiaxes is independent of the radius:
\be
\frac{\xi_{\phi}^{\star}}{\xi_{\theta}^{\star}}=-\frac{4lk}{5\varepsilon_{\sR}}
\sin^2\theta\cos\theta\frac{F_2}{\frac{m}{\varepsilon_{\sR}}F_1-\frac{4lk}{5}
 F_2 \sin^2\theta} .
\ee
It is seen that near the equator and near the pole the motion is mainly in
$\vec{e_{\theta}}$ direction. For large $l \gg 1$ this ratio is
\be
\frac{\xi_{\phi}^{\star}}{\xi_{\theta}^{\star}} \simeq 
\frac{14k}{5\varepsilon_{\sR}}\, \frac{\sin\theta\tan\lambda}
{\frac{m}{\varepsilon_{\sR}}+\frac{14}{5}k\tan\theta\tan\lambda} .
\ee
The flow pattern considered here drifts opposite to the sense of
rotation with a speed $v_{\rm ph}$, if it is observed in a frame rotating
with the star.

\section{Conclusions}

In the present work we have derived the general PDE governing non-radial,
adiabatic, long-period (with respect to the rotation period), linear
oscillations of a slowly and differentially rotating star. This general
equation includes all the high-order $g$-modes and all possible hybrids of
rotation modes as well as their mutual interaction. The geophysical 
`traditional approximation' considerably simplifies this general equation,
and we get two ODEs for the $r$- and $\theta$-components instead of one with
arbitrary gradients of rotation $\Omega(r,\theta)$. We have received a more
stringent condition for the applicability of this approximation to the
pulsation of stars. Only for very low frequencies this restriction is the
same as that of the standard case.

The $\theta$-equation is Laplace's equation generalized to the latitudinal
differential rotation. Without solving this equation qualitatively we found
the exact condition for the appearance of a global instability. This 
instability is driven by the latitudinal shear, it is not influenced by 
buoyancy. We call that a `latitudinal Kelvin-Helmholtz instability'
(LKHI). The appearance of LKHI strongly depends on the Rossby number (the
ratio of rotation period and period of motion), on the azimuthal wave numbers
and on the latitudinal rotation gradients. Very large gradients produce
retrograde waves (seen in the rotating frame), while a slower rotation
gradient is responsible for prograde mode LKHI. The rotation gradient has
a lower boundary below which LKHI is not possible for any Rossby number
or azimuthal number $m$.

We have applied the LKHI condition to the helioseismological data of the
Sun. Here a global LKHI is possible for the $m=1$ mode at practically
all latitudes. Radially the LKHI is extended from the greatest part of the
tachocline up to the photosphere. The LKHI for the Sun was first obtained
by Watson (1981). According to his results the instability is possible only
at photosphere layers. Later Gilman \& Fox (1997) have shown that such an
instability is possible in the tachocline too, if strong toroidal magnetic
fields are included. Our results show that the instability of the $m=1$
modes and other modes is possible without magnetic fields, in contradiction
to Gilman \& Fox (1997). This difference is probably connected with the
incompleteness of the equations used by Watson (1981) and by Gilman \& Fox
(1997); their equations are two-dimensional only.

The exact solutions of Laplace's tidal equation for lower frequencies are
expressed by Jacobi's polynomials. Just for lower frequencies the numerical
calculations of stellar pulsation analyses meet great problems, when looking
for the eigenfunction as infinite series of Legendre functions. The
eigenfunctions, defined by higher-order polynomials of Jacobi, cannot be
expressed by convergent series of associated Legendre functions. Every
Legendre function is a particular case of a Jacobi polynomial.

It has been shown here that the retrograde (slow and fast) modes with high 
surface wave numbers ($l,m$) are energetically concentrated in narrow bands
of latitudes. This analysis was done for the 22-year modes as an example.
Such a concentration of mode energy in a narrow spatial area makes such modes
vulnerable to different instability mechanisms such as the $\en$--mechanism
considered in Paper~1.  

\appendix

\begin{table*} 
 \centering
 \caption{Results for the 22-year period: frequency deviations
 $\delta=\omega_{22}-\omega$ for permitted quantum numbers $(l, |m|)$}  
 \begin{tabular}[]{rrrrr|@{\ \ }|rrrrr|@{\ \ }|rrrrr}
    \hline
  $l$  & $m$ \ & $\delta\ \ \ $ &$m$ \ & $\delta\ $ & 
    $l$  & $m$ \ & $\delta\ \ \ $ &$m$ \ & $\delta\ $ &$l$ & $m$ \ &
  $\delta\ \ \ $ & $m$ \ & $\delta\ \ \ $ \\
  & fast& (nHz) &  slow & (nHz)  &   & fast& (nHz) &  slow & (nHz)  & 
    &  fast & (nHz)  & slow & (nHz) \\
    \hline\hline                           
 11 &   1 &  0.055 & 575 & 0.000 & 
 33 &  10 & -0.021 & 478 & 0.000 &
 55 &  35 &-0.008 &   365 &-0.001\\
 12 &   1 & 0.250  & 571 & 0.001 & 
 34 &  10 & 0.051  & 474 & 0.001 &
 56 &  36 & 0.010  & 360 & 0.001 \\
 13 &   1 & 0.406 & 567 & 0.001 &
 35 &  11 & 0.020 & 469 & 0.000 & 
 57 &  38 & 0.007  & 354 & 0.001 \\  
 14 &   2 &-0.265 & 562 &-0.001 &  
 36 &  12 &-0.003 & 464 & 0.000 &    
 58 &  40 & 0.007 & 348 & 0.001 \\
 15 &   2 &-0.073 & 558 & 0.000 &   
 37 &  13 &-0.021 & 459 &-0.001 &  
 59 &  43 &-0.008 & 341 &-0.001 \\
 16 &   2 & 0.089 & 554 & 0.000 &  
 38 &  14 &-0.033 & 454 &-0.001 &   
 60 &  45 &-0.004 & 335 & 0.000 \\
 17 &   2 & 0.226 & 550 & 0.001 &  
 39 &  14 & 0.029 & 450 & 0.001 &     
 61 &  47 & 0.002 & 329 & 0.000 \\ 
 18 &   3 &-0.128 & 545 &-0.001 & 
 40 &  15 & 0.020 & 445 & 0.001 & 
 62 &  50 &-0.005 & 322 &-0.001 \\  
 19 &   3 & 0.013 & 541 & 0.000 &
 41 &  16 & 0.014 & 440 & 0.001 &  
 63 &  52 & 0.004 & 316 & 0.001 \\ 
 20 &   3 & 0.136 & 537 & 0.001 &  
 42 &  17 & 0.012 & 435 & 0.000 & 
 64 &  55 & 0.001 & 309 & 0.000 \\
 21 &   4 &-0.091 & 532 &-0.001 &
 43 &  18 & 0.012 & 430 & 0.000 & 
 65 &  58 & 0.002 & 302 & 0.000 \\
 22 &   4 & 0.028 & 528 & 0.000 &
 44 &  19 & 0.014 & 425 & 0.001 &
 66 &  61 & 0.004 & 295 & 0.001 \\ 
 23 &   4 & 0.134 & 524 & 0.001 &    
 45 &  20 & 0.018 & 420 & 0.001 &
 67 &  65 &-0.001 & 287 & 0.000 \\ 
 24 &   5 &-0.021 & 519 & 0.000 &  
 46 &  22 &-0.019 & 414 &-0.001 &
 68 &  69 &-0.002 & 279 &-0.001 \\ 
 25 &   5 & 0.079 & 515 & 0.001 &  
 47 &  23 &-0.010 & 409 &-0.001 &
 69 &  73 &-0.001 & 271 & 0.000 \\ 
 26 &   6 &-0.035 & 510 & 0.000 &   
 48 &  24 & 0.001 & 404 & 0.000 &
 70 &  77 & 0.002 & 263 & 0.000 \\
 27 &   6 & 0.058 & 506 & 0.001 &   
 49 &  25 & 0.011 & 399 & 0.001 &
 71 &  82 & 0.001 & 254 & 0.000 \\
 28 &   7 &-0.029 & 501 & 0.000 &
 50 &  27 &-0.009 & 393 &-0.001 &
 72 &  88 & 0.000 & 244 & 0.000 \\ 
 29 &  7  & 0.057 & 497 & 0.001 &  
 51 &  28 & 0.005 & 388 & 0.000 &
 73 &  94 & 0.001 & 234 & 0.001 \\ 
 30 &   8 &-0.009 & 492 & 0.000 &  
 52 &  30 &-0.009 & 382 &-0.001 &
 74 & 102 & 0.000 & 222 & 0.000 \\
 31 &   9 &-0.060 & 487 &-0.001 &  
 53 &  31 & 0.006 & 377 & 0.001 &
 75 & 112 &-0.001 & 208 & 0.000 \\
 32 &   9 & 0.019 & 483 & 0.000 &
 54 &  33 &-0.002 & 371 & 0.000 & 
 76 & 125 & 0.000 & 191 & 0.000 \\ 
\hline  
\end{tabular}
\end{table*}    

\section{Coefficients of the main equation}

In Section~2.2 our main Eq.~(26) for the pressure perturbations has been
derived:
\be
\left[\psi_1\check{\delta}^2_r + \psi_2\check{\delta}^2_{\mu} +
\psi_3\check{\delta}_r + \psi_4\check{\delta}_{\mu} +
\psi_5\check{\delta}_r\check{\delta}_{\mu} + \psi_6\right]\tilde{P} = 0 .
\nonumber
\ee
This singular PDE has the following coefficients:
\noindent
\begin{eqnarray*}
\psi_1 &=& 1-\frac{N^2}{\omega^2} - j\mu^2\frac{1+\beta_r}{\alpha} ,\\
\psi_2 &=& \psi_1\frac{1-\mu^2}{\alpha}\left[\frac{\epsilon^2_R}{\mu^2}\left(1-
\frac{N^2}{\omega^2}\right) - j(1+\beta_r)\right] ,\\ 
\psi_3 &=& f_1 + \psi_1f_2 ,\\
\psi_4 &=& \frac{\psi_1^2}{\mu^2}\left[\frac{a_7}{a_3} + 
 (1-\mu^2)\left(\check{\delta}_{\mu}\frac{1}{a_3}-
\frac{a_5}{a_3}\right)\right] +\\
&+& j\left[a_9\psi_1(f_2-\check{\delta}_{\mu}a_3^{\ast}) -
f_1 a_3^{\ast} - \psi_1\check{\delta}_r a_3^{\ast}\right] ,\\
\psi_5 &=& j\psi_1\frac{1-\mu^2}{\alpha}\left(2+\beta_r+\beta_{\mu}\right) ,\\
\psi_6 &=& f_1 f_2 + \psi_1(\check{\delta}_r f_2) + 
 j\psi_1a_9(\check{\delta}_{\mu} f_2) - \\
&-& \frac{\psi_1^2}{\mu^2}\left[m^2+\frac{a_7 a_5}{a_3}+
(1-\mu^2)\check{\delta}_{\mu}\left(\frac{a_5}{a_3}\right)\right] ,
\end{eqnarray*}
where
\begin{eqnarray*}
f_1 &=& \psi_1(2+j a_8)-\check{\delta}_r \psi_1 - j a_9\check{\delta}_{\mu}\psi_1 ,\\
f_2 &=& 1 + \kap_{\rho} - \frac{m}{\varepsilon_{\sR}}\beta_r - j a_6 ,\\
a_3^{\ast} &=& (\mu^2 -1)(1+\beta_{\mu})/\alpha=1-1/a_3 .
\end{eqnarray*}
Taking into account $\nabla\omega = -m\nabla\Omega$ we can easily obtain all
the required derivations of the parameters.

\begin{acknowledgements}
   
The authors gratefully acknowledge financial support of the present work by
the German Science Foundation (DFG) under grant No. 436 RUS 113/560/4-1.
\end{acknowledgements}


{}

\end{document}